\documentclass[longauth]{aa}  
\usepackage[varg]{txfonts}
\usepackage{graphicx}
\usepackage{lscape}
\usepackage[breaklinks,colorlinks,citecolor=blue]{hyperref}
\usepackage{natbib}
\bibpunct{(}{)}{;}{a}{}{,} % to follow the A&A style
\begin{document}

\title{The CARMENES search for exoplanets around M dwarfs}  
\subtitle{Rotational variation in activity indicators of Ross~318, YZ~CMi,\\TYC~3529-1437-1, and EV~Lac}
\titlerunning{Rotational variation in activity indicators of four stars}
\author{P.~Sch{\"o}fer\inst{\ref{IAA},\ref{IAG}}
        \and
        S.~V.~Jeffers\inst{\ref{MPS}}
        \and
        A.~Reiners\inst{\ref{IAG}}
        \and
        M.~Zechmeister\inst{\ref{IAG}}
        \and
        B.~Fuhrmeister\inst{\ref{HS}}
        \and
        M.~Lafarga\inst{\ref{ICE1},\ref{ICE2},\ref{Warwick}}
        \and
        I.~Ribas\inst{\ref{ICE1},\ref{ICE2}}
        \and
        A.~Quirrenbach\inst{\ref{LSW}}
        \and
        P.~J.~Amado\inst{\ref{IAA}}
        \and
        J.~A.~Caballero\inst{\ref{CAB}}
        \and
        G.~Anglada-Escud{\'e}\inst{\ref{IAA},\ref{QMUL}}
        \and
        F.~F.~Bauer\inst{\ref{IAA}}
        \and
        V.~J.~S.~B{\'e}jar\inst{\ref{IAC1},\ref{IAC2}}
        \and
        M.~Cort{\'e}s-Contreras\inst{\ref{CAB}}
        \and
        E.~D{\'{\i}}ez~Alonso\inst{\ref{UCM},\ref{Oviedo}}
        \and
        S.~Dreizler\inst{\ref{IAG}}
        \and
        E.~W.~Guenther\inst{\ref{TLS}}
        \and
        O.~Herbort\inst{\ref{Graz},\ref{StAnd1},\ref{StAnd2},\ref{StAnd3}}
        \and
        E.~N.~Johnson\inst{\ref{IAG}}
        \and
        A.~Kaminski\inst{\ref{LSW}}
        \and
        M.~K{\"u}rster\inst{\ref{MPIA}}
        \and
        D.~Montes\inst{\ref{UCM}}
        \and
        J.~C.~Morales\inst{\ref{ICE1},\ref{ICE2}}
        \and
        S.~Pedraz\inst{\ref{CAHA}}
        \and
        L.~Tal-Or\inst{\ref{TAU},\ref{IAG}}
        }
\institute{
        Instituto de Astrof\'{\i}sica de Andaluc\'{\i}a (IAA-CSIC), Glorieta de la Astronom\'{\i}a s/n, E-18008, Granada, Spain\\\email{pschoefer.astro@freenet.de}\label{IAA}
        \and
        Institut f{\"u}r Astrophysik, Friedrich-Hund-Platz 1, D-37077 G{\"o}ttingen, Germany\label{IAG}
        \and
        Max Planck Institute for Solar System Research, Justus-von-Liebig-Weg 3, D-37077 G{\"o}ttingen, Germany\label{MPS}
        \and
        Hamburger Sternwarte, Gojenbergsweg 112, D-21029 Hamburg, Germany\label{HS}
        \and
        Institut de Ci{\`e}ncies de l'Espai (ICE, CSIC), Campus UAB, c/ de Can Magrans s/n, E-08193, Bellaterra, Barcelona, Spain\label{ICE1}
        \and
        Institut d'Estudis Espacials de Catalunya (IEEC), C/ Gran Capit\`a 2-4, E-08034 Barcelona, Spain\label{ICE2}
        \and
        Department of Physics, University of Warwick, Gibbet Hill Road, Coventry CV4 7AL, United Kingdom\label{Warwick}
        \and
        Landessternwarte, Zentrum f{\"u}r Astronomie der Universit{\"a}t Heidelberg, K{\"o}nigstuhl 12, D-69117, Heidelberg, Germany\label{LSW}
        \and
        Centro de Astrobiolog\'{\i}a (CSIC-INTA), ESAC, Camino Bajo del Castillo s/n, E-28692, Villanueva de la Ca\~{n}ada, Madrid, Spain\label{CAB}
        \and
        School of Physics and Astronomy, Queen Mary, University of London, 327 Mile End Road, London, E1 4NS, UK\label{QMUL}
        \and
        Instituto de Astrof\'{\i}sica de Canarias, V\'{\i}a L{\'a}ctea s/n, E-38205 La Laguna, Tenerife, Spain\label{IAC1}
        \and
        Departamento de Astrof\'{\i}sica, Universidad de La Laguna, E-38206 La Laguna, Tenerife, Spain\label{IAC2}
        \and
        Departamento de F\'{\i}sica de la Tierra y Astrof\'{\i}sica \& IPARCOS-UCM (Instituto de F\'{\i}sica de Part\'{\i}culas y del Cosmos de la UCM), Facultad de Ciencias F\'{\i}sicas, Universidad Complutense de Madrid, E-28040 Madrid, Spain\label{UCM}
        \and
        Departamento de Explotaci{\'o}n y Prospecci{\'o}n de Minas, Escuela de Minas, Energ\'{\i}a y Materiales, Universidad de Oviedo, E-33003 Oviedo, Asturias, Spain\label{Oviedo}
        \and
        Th{\"u}ringer Landessternwarte Tautenburg, Sternwarte 5, D-07778 Tautenburg, Germany\label{TLS}
        \and
        Space Research Institute, Austrian Academy of Sciences, Schmiedlstrasse 6, A-8042 Graz, Austria\label{Graz}
        \and
        Centre for Exoplanet Science, University of St Andrews, North Haugh, St Andrews, KY16 9SS, UK\label{StAnd1}
        \and
        SUPA, School of Physics \& Astronomy, University of St Andrews, North Haugh, St Andrews, KY16 9SS, UK\label{StAnd2}
        \and
        School of Earth \& Environmental Sciences, University of St Andrews, Irvine Building, St Andrews, KY16 9AL, UK\label{StAnd3}
        \and
        Max-Planck-Institut f{\"u}r Astronomie, K{\"o}nigstuhl 17, D-69117, Heidelberg, Germany\label{MPIA}
        \and
        Centro Astron{\'o}mico Hispano-Alem{\'a}n (CSIC-MPG), Observatorio Astron{\'o}mico de Calar Alto, Sierra de los Filabres, E-04550 G{\'e}rgal, Almer\'{\i}a, Spain\label{CAHA}
        \and
        Department of Physics, Ariel University, Ariel 40700, Israel\label{TAU}
        }
\date{Received 17 October 2019 / Accepted 7 April 2022}

% \abstract{}{}{}{}{} 
% 5 {} token are mandatory
\abstract
% context heading (optional), leave it empty if necessary 
{The Calar Alto high-Resolution search for M dwarfs with Exo-earths with Near-infrared and optical \'Echelle Spectrographs (CARMENES) instrument is searching for periodic radial-velocity (RV) variations of M dwarfs, which are induced by orbiting planets. However, there are other potential sources of such variations, including rotational modulation caused by stellar activity.}
{We aim to investigate four M dwarfs (Ross~318, YZ~CMi, TYC~3529-1437-1, and EV~Lac) with different activity levels and spectral sub-types. Our goal is to compare the periodicities seen in 22 activity indicators and the stellar RVs, and to examine their stability over time.}
{For each star, we calculated generalised Lomb-Scargle periodograms of pseudo-equivalent widths of chromospheric lines, indices of photospheric bands, the differential line width as a measure of the width of the average photospheric absorption line, the RV, the chromatic index that describes the wavelength dependence of the RV, and parameters of the cross-correlation function. We also calculated periodograms for subsets of the data and compared our results to TESS photometry.}
{We find the rotation periods of all four stars to manifest themselves in the RV and photospheric indicators, particularly the TiO~7050 index, whereas the chromospheric lines show clear signals only at lower activity levels. For EV~Lac and TYC~3529-1437-1, we find episodes during which indicators vary with the rotation period, and episodes during which they vary with half the rotation period, similarly to photometric light curves.}
{The changing periodicities reflect the evolution of stellar activity features on the stellar surface. We therefore conclude that our results not only emphasise the importance of carefully analysing indicators complementary to the RV in RV surveys, but they also suggest that it is also useful to search for signals in activity indicators in subsets of the dataset, because an activity signal that is present in the RV may not be visible in the activity indicators all the time, in particular for the most active stars.}
\maketitle
%
%________________________________________________________________

\section{Introduction}
\label{section.introduction}
Current radial velocity (RV) surveys such as the Calar Alto high-Resolution search for M dwarfs with Exo-earths with Near-infrared and optical \'Echelle Spectrographs \citep[CARMENES;][]{2018SPIE10702E..0WQ} are searching for exoplanets that reveal themselves through periodic RV variations of their host stars down to $1\,\mathrm{m\,s}^{-1}$ or less. The RV is derived from the observed wavelengths of spectral lines and can therefore be affected by line profile changes caused by star spots or plages. In consequence, inhomogeneous distributions of stellar surface features that lead to variations over the course of one rotation, the evolution of active regions, and activity cycles can introduce quasi-periodic RV signals of the same order of magnitude as prospective planetary signals \citep[e.g.][]{1997ApJ...485..319S,2013ApJ...770..133H,2014MNRAS.439.3094B,2015ApJ...812...42B}.

To identify activity-induced RV signals, the variations on the stellar surface need to be monitored using indicators that are complementary to the RV. Both a photometric approach using brightness variations induced by active regions \citep[e.g.][]{2007AcA....57..149K,2011ApJ...727...56I,2016ApJ...821...93N,2016A&A...595A..12S,2019A&A...621A.126D} and a spectroscopic approach based on variations of chromospheric emission lines such as \ion{Ca}{ii}~H\&K and H$\alpha$ \citep[e.g.][]{1984ApJ...279..763N,2015MNRAS.452.2745S,2018A&A...612A..89S,2019A&A...623A..24F} have been used to uncover rotation periods and long-term activity cycles of late-type stars. Over time, evolution and migration of active regions change their distribution on the stellar surface. This fact increases the difficulty of identifying the rotation period, as, for example, photometric light curves often show one dip per rotation for some time, but two dips at other times \citep{2018ApJ...863..190B}. Since active regions may predominantly appear at different latitudes at different times, the detected period may also be subject to variations caused by differential rotation \citep{1996ApJ...466..384D}. More recent works show that other effects can impact the light curve in similar ways to differential rotation \citep[e.g.][]{2020ApJ...901...14B}.

An activity study of 331 M dwarfs of the CARMENES sample was presented in \citet{2019A&A...623A..44S}. That work explored the behaviour of eight chromospheric activity indicators and four photospheric absorption band indices for different spectral subtypes and the correlations among these indicators. In addition, three best-fit periodicities for each indicator were derived from generalised Lomb-Scargle (GLS) periodograms \citep{2009A&A...496..577Z} and compared to photometric rotation periods from the literature. This revealed that the photospheric titanium oxide $\lambda7050\,${\AA} absorption band head, the middle line of the \ion{Ca}{ii} infrared triplet, and the chromospheric H$\alpha$ emission line most commonly show rotational variation, and that 11\% of the stars with previously known, mostly photometrically derived rotation periods, appear to show a rotational signal in more than two activity indicators. \citet{2021A&A...652A..28L} studied rotational signals in a different set of ten spectroscopic indicators in the visible-light range for a subset of 98 CARMENES sample stars and found that any given indicator shows a signal most commonly in a specific mass and activity regime. They also indentified five stars that are representative of the types of rotational signals seen in their sample. A detailed study of periodicities and also correlations between spectroscopic indicators was carried out by \citet{2022arXiv220300415J} for the very active M dwarf EV~Lac.

In this work, we analysed the rotational variation of 24 spectroscopic indicators in both the visible-light and the near-infrared range in more detail, including the RV, for four M dwarfs that show particularly clear signals. Ross~318, YZ~CMi, and EV~Lac are representative examples for the different groups defined by \citet{2021A&A...652A..28L}, while TYC~3529-1437-1 switched groups over time. We describe the target stars in Sect.~\ref{section.targets}, and our data and indicators in Sect.~\ref{section.observations}. The method for our periodicity analysis is described in Sect.~\ref{section.analysis}, the results are presented and discussed in Sects.~\ref{section.spec} and \ref{section.phot}. Finally, in Sect.~\ref{section.conclusions} we summarise our main conclusions.

%-------------------------------------------------------------------------------------------------------------------------------------------
\section{Target stars}
\label{section.targets}
\begin{figure}
  \resizebox{\hsize}{!}{\includegraphics{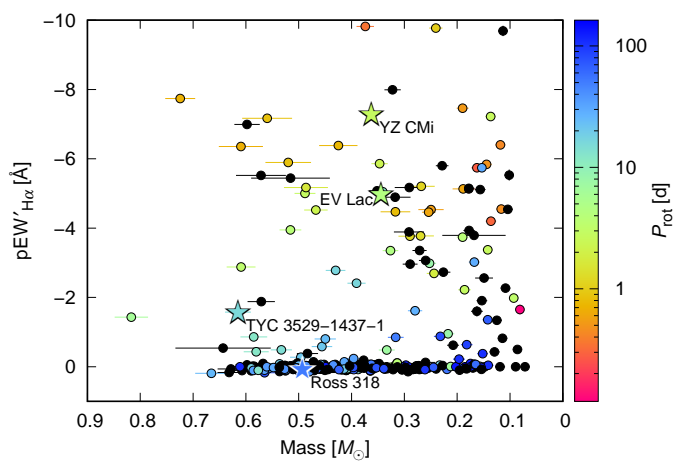}}
  \caption{Pseudo-equivalent width of H$\alpha$ line after subtracting an inactive reference star spectrum (pEW$'_{\mathrm{H}\alpha}$) as a function of the stellar mass for 331 M dwarfs in the CARMENES sample, colour-coded by the rotation period $P_\mathrm{rot}$. Black points are stars with unknown $P_\mathrm{rot}$. The four stars studied in this work are marked with star symbols and their names. pEW$'_{\mathrm{H}\alpha}$ and $P_\mathrm{rot}$ are adopted from \citet{2019A&A...623A..44S} and references therein, stellar masses were generally computed from luminosities and effective temperatures as in \citet{2019A&A...625A..68S} and \citet{2020A&A...642A.115C}, but with the latest Gaia EDR3 parallax and photometry \citep{2021A&A...649A...1G}. More details on these updated masses will be provided in a forthcoming publication.}
  \label{fig:pEW_mass}
\end{figure}

The CARMENES survey focuses on the brightest M dwarfs of each spectral sub-type from M0.0\,V to M9.0\,V that are observable from the Calar Alto Observatory. Since the start of the survey in January 2016, a sample of more than 380 stars has been observed \citep{2019A&A...625A..68S,2021A&A...653A.114S}. While known binaries with a separation of less than 5\,arcsec were excluded, there was no target pre-selection based on further criteria such as stellar activity and metallicity. Therefore, the CARMENES sample contains wide ranges of stellar parameters such as mass, rotation, and chromospheric activity level, as is shown in Fig.~\ref{fig:pEW_mass}.

\citet{2019A&A...623A..44S} reported that 15 out of 133 CARMENES sample stars with a rotation period $P_\mathrm{rot} > 1\,$d known at that date showed this period as one of the three best-fit periodicities in the variations of three or more chromospheric or photospheric activity indicators. However, the periodograms for 11 of these stars generally showed a forest of peaks, and the frequency corresponding to the rotation period was only the second or third strongest signal in the periodogram. It is possible that the strongest signals were introduced by the evolution of active regions similarly to the RV signals in \citet{2020AJ....159...23N}. Several of these stars were observed fewer than 40 times, and thus did not ensure a sufficient sampling rate for a detailed period analysis. In contrast, the four remaining stars were already observed more than 40 times and the rotation period clearly appeared as the most significant frequency in more than two activity indicators. In this work, we studied the rotational variation in the activity indicators of these four stars in more detail. The four stars are highlighted in Fig.~\ref{fig:pEW_mass} and listed in Table~\ref{table:targets}.

\begin{table*}
\caption{Basic parameters, amplitude of periodic RV variations, $A_\mathrm{RV}$, and number of CARMENES observations, $N_\mathrm{obs}$, of target stars.}
\label{table:targets}
\centering
\small
\begin{tabular}{ll cccccccc}
\hline\hline
\noalign{\smallskip}
Karmn & Name & $\alpha$ & $\delta$  & Spectral & $P_\mathrm{rot}$ & $v\sin i$ & $\log(L_{\mathrm{H}\alpha}/L_\mathrm{bol})$ & $A_\mathrm{RV}$ & $N_\mathrm{obs}$\\
 & & [hh:mm:ss.ss] & [dd:mm:ss.s] & type & [d] & [km\,s$^{-1}$] & & [m\,s$^{-1}$] & \\
\hline
\noalign{\smallskip}
J01025+716 & Ross 318 & 01:02:32.24 & +71:40:47.3 & M3.0\,V & $51.5\pm 2.6$ & $< 2$ & \ldots & 3 & 120\\
J07446+035 & YZ CMi & 07:44:40.17 & +03:33:08.9 & M4.5\,V & $2.78\pm 0.01$ & $4.0\pm 1.5$ & $-3.61\pm 0.01$ & 57 & 51\\
J18174+483 & TYC 3529-1437-1 & 18:17:25.13 & +48:22:02.3 & M2.0\,V & $15.8\pm 0.1$ & $< 2$ & $-4.03\pm 0.01$ & 11 & 71\\
J22468+443 & EV Lac & 22:46:49.73 & +44:20:02.4 & M3.5\,V & $4.38\pm 0.03$ & $3.5\pm 1.5$ & $-3.65\pm 0.01$ & 39 & 107\\
\hline
\end{tabular}
\tablefoot{Carmencita identifiers as defined by \citet{2015A&A...577A.128A}, Simbad names, equatorial coordinates (J2000.0) from Gaia EDR3 \citep{2021A&A...649A...1G}, spectral types from \citet{1995AJ....110.1838R} and \citet{2006AJ....132..866R}, rotation periods from \citet{2019A&A...621A.126D}, projected rotational velocities from \citet{2018A&A...612A..49R}, and normalised H$\alpha$ luminosities from \citet{2019A&A...623A..44S}.}
\end{table*}

Three of our target stars are representative examples for the groups of different rotational signals defined in \citet{2021A&A...652A..28L}: Ross~318 has a long rotation period above $50\,$d, for which signals are hard to detect, particularly in chromospheric emission lines \citep{2019A&A...623A..24F}, YZ~CMi was the example for a star with clear signals at the rotation period, and EV~Lac was the example for multiple signals related to the rotation period. TYC~3529-1437-1 showed clear signals at the rotation period in \citet{2019A&A...623A..44S}, but signals at multiple related frequencies in \citet{2021A&A...652A..28L}. As we selected the stars with the most significant signals at known rotation periods, this work does not include examples for signals at a previously unknown rotation period or with insignificant signals.

Despite the small sample size, our target stars cover a diverse parameter space. Ross~318 is an inactive mid-type M~dwarf with no H$\alpha$ emission, as suggested by its long rotation period, and has neither a large RV scatter nor a detectable projected rotational velocity. TYC~3529-1437-1 is an early-type M dwarf with a significantly shorter rotation period of $15.8\,$d and moderate chromospheric activity. In addition, it shows a large, periodically modulated RV scatter with amplitude of $A_\mathrm{RV}=11\,\mathrm{m\,s}^{-1}$, while its projected rotational velocity is below the detection limit of $2\,\mathrm{km\,s}^{-1}$ \citep{2018A&A...612A..49R}. By contrast, the mid-type M~dwarfs YZ~CMi and EV~Lac both show large RV scatters and a detectable projected rotational velocity, and they are therefore part of the CARMENES sample of active RV-loud stars \citep{2018A&A...614A.122T}. Both stars are among the most active stars in the CARMENES sample with normalised H$\alpha$ luminosities $\log(L_{\mathrm{H}\alpha}/L_\mathrm{bol}) \approx -3.6$ and rotation periods below $5\,$d. For YZ~CMi, \citet{2020A&A...641A..69B} used CARMENES spectra and photometric data to constrain star-spot coverage and temperatures and to find a convective redshift.

The rotation periods were derived by \citet{2019A&A...621A.126D} using photometric time series from the ASAS-SN (Ross~318), MEarth (YZ~CMi), and SuperWASP (TYC~3529-1437-1, EV~Lac) surveys. Their results are in agreement with previously reported values of $2.78\,$d for YZ~CMi \citep{1974IzKry..52....3C} and $4.376\,$d for EV~Lac \citep{1995A&A...300..819C}, and the more recently measured value of $49.4642\,$d for Ross~318 \citep{2020MNRAS.491.5216G} from independent photometric data. The value of $16.2578\,$d was also found for TYC~3529-1437-1 by \citet{2007A&A...467..785N} in an earlier subset of the SuperWASP data. This indicates a high reliability of the rotation periods for these stars.

%-------------------------------------------------------------------------------------------------------------------------------------------
\section{Observations}
\label{section.observations}
In this section, we describe the high-resolution spectroscopic CARMENES data and the photometric data obtained with the TESS satellite.

\subsection{CARMENES}
\label{subsection.carmenes}
We analysed a total of 349 CARMENES Guaranteed Time Observations collected between 3 January 2016 and 17 January 2019. The number of observations for each of the four stars is given in Table~\ref{table:targets}. For all observations, the visible-light (VIS) and near-infrared (NIR) channels of CARMENES were used simultaneously. After spectral reduction with the \texttt{CARACAL} pipeline \citep{2016SPIE.9910E..0EC} using flat-relative optimal extraction \citep{2014A&A...561A..59Z}, we derived 24 spectroscopic indicators from the spectra. These include pseudo-equivalent widths after subtracting an inactive reference star spectrum of the same spectral type \citep[pEW$'$,][]{2019A&A...623A..44S} of spectral lines with a chromospheric component, including \ion{He}{i}~D$_3$, \ion{Na}{i}~D doublet, H$\alpha$, and the \ion{Ca}{ii} infrared-triplet lines (IRT-a, -b, and -c from shortest to longest wavelength) in the VIS channel, and \ion{He}{i}~$\lambda$10833\,{\AA} and Pa$\beta$ in the NIR channel. Next, indices of photospheric absorption bands, including TiO~7050, TiO~8430, VO~7436, and VO~7942 \citep{2019A&A...623A..44S} are present. We also derived a differential line width \citep[dLW,][]{2018A&A...609A..12Z} in both the VIS and the NIR channel, as well as radial velocity (RV~VIS and RV~NIR) in both spectrograph channels calculated using \texttt{SERVAL} \citep{2018A&A...609A..12Z} with a correction for an instrumental nightly zero-point offset, as described by \citet{2018A&A...609A.117T} and \citet{2019MNRAS.484L...8T}, and the corresponding chromatic indices \citep[CRX~VIS and CRX~NIR,][]{2018A&A...609A..12Z}, which measure the wavelength dependence of the RV. Finally, we included a full width at half minimum (FWHM), contrast, and bisector inverse slope \citep[BIS;][]{2001A&A...379..279Q} of the cross-correlation function (CCF) with a weighted binary mask from co-added spectra of the star for both spectrograph channels as described in \citet{2020A&A...636A..36L}.

While the same VIS channel spectra were also used in the study of rotational signals in a larger sample of stars by \citet{2021A&A...652A..28L}, we extended the set of indicators by including the NIR channel spectra and the photospheric absorption band indices and used a different measure for the chromospheric lines.

\subsection{TESS}
\label{subsection.tess}
All four stars were also observed by the Transiting Exoplanet Survey Satellite \citep[TESS,][]{2015JATIS...1a4003R}: Ross~318 in observing sectors 18, 19, 24, and 25; YZ~CMi in sectors 7 and 34; TYC~3529-1437-1 in sectors 14, 25, 26, 40, and 41; and EV~Lac in sector 16. While these photometric observations are not contemporaneous with the CARMENES observations, we still use them for comparison.

We retrieved the TESS light curves from the Mikulski Archive for Space Telescopes\footnote{\url{https://mast.stsci.edu/}} and used the simple aperture photometry (SAP) flux. For TYC~3529-1437-1 in observing sector 14 and YZ~CMi in sector 34, we additionally used the regression corrector from \texttt{Lightkurve} \citep{2018ascl.soft12013L} to remove instrument systematics that dominated the SAP flux.

%-------------------------------------------------------------------------------------------------------------------------------------------
\section{Analysis}
\label{section.analysis}
We used GLS periodograms \citep{2009A&A...496..577Z} to reveal periodic changes in all our measured indicators. To mitigate the impact of flaring events or bad spectra, we rejected any values that deviate from the respective average values by more than twice the standard deviation (2$\sigma$ clipping). We also rejected \ion{He}{i}~$\lambda$10833\,{\AA} and Pa$\beta$ measurements of spectra with an observed RV that shifts the neighbouring strong telluric lines into the respective line windows. In the case of YZ~CMi, this reduced the number of used measurements of these two indicators by more than 50\%.

Because the stars were usually observed at most once per night and all four stars have rotation periods $P_\mathrm{rot} = 1/f_\mathrm{rot}$ longer than one day, we limited the periodograms to frequencies up to $1\,\mathrm{d}^{-1}$. The nightly sampling introduces alias signals at frequencies $f' = 1\,\mathrm{d}^{-1}-f$ for any signal at frequency $f$. For the two slower rotating stars, we therefore only show the periodograms (Figs.~\ref{fig:Ross318_gls} and \ref{fig:TYC_gls}) up to $0.5\,\mathrm{d}^{-1}$ for clarity in the relevant frequency range because the signals at higher frequencies are likely alias signals. We quantified the significance of a signal at frequency $f$ by calculating the probability $p(f)$ that a power higher than the observed GLS power at $f$ was produced by Gaussian noise. We considered a signal significant if its significance exceeds $3\sigma$, that is, $\log p(f) < -2.87$.

In addition to these overall GLS periodograms of the full dataset for each star, we calculated periodograms of subsets of consecutive data points to test the stability of the signals in the overall periodogram and to investigate whether the most significant signals appear at the same frequencies all the time. Our datasets are not evenly sampled in time, but they contain gaps of up to several months, during which the star is not well observable from Calar Alto. Therefore, we split the dataset for each star into two subsets, one before and one after a significant sampling gap. In the case of EV~Lac, we further split the subset after the sampling gap into two parts, the second of which contains observations with a higher cadence than the first. Our subsets for EV~Lac include, but are not limited to the observations used for the low-resolution Doppler imaging maps 1--4, map 5, and maps 6--8 in the analysis by \citet{2022arXiv220300415J}, respectively. A group of six spectra of YZ~CMi between October 2017 and January 2018 with a large gap from the preceding subset is too small for further analysis and thus excluded. Similarly, a single isolated spectrum of EV~Lac in January 2016 is excluded from further analysis.

%-------------------------------------------------------------------------------------------------------------------------------------------
\section{Spectroscopic periodicities and their stability and evolution}
\label{section.spec}
In this section, we present and discuss the results of our periodogram analysis described in the previous section. For each star, we start with the rotational signals in the full datasets and then assess their stability and evolution using the data subsets.

\subsection{Ross~318}
\label{subsection.Ross318}
We show the GLS periodograms of all our indicators for the slow-rotating ($P_\mathrm{rot}=(51.5\pm 2.6)\,$d) M3.0\,V star \object{Ross~318} in Fig.~\ref{fig:Ross318_gls} and tabulate the mean values, standard deviations, and probabilities $\log p(f_\mathrm{rot})$ for each indicator in Table~\ref{table:Ross318}. As reported by \citet{2019A&A...623A..44S} and \citet{2019A&A...623A..24F}, the chromospheric lines show only small variations in the least active M dwarfs, and, in consequence, solid detections of $P_\mathrm{rot} \gtrsim 50\,$d using these indicators are elusive. Still, we find that H$\alpha$, two of the \ion{Ca}{ii} IRT lines, and \ion{He}{i}~$\lambda$10833\,{\AA} show the highest periodogram peak with a significance of more than $3\sigma$ at $f_\mathrm{rot}$, while \ion{Ca}{ii} IRT-a and Pa$\beta$ show significant power but not the highest peak within the 3$\sigma$ confidence interval of the rotation frequency.

TiO~7050, CRX~VIS, and CCF~FWHM in both spectrograph channels all show a significant highest periodogram peak at $f_\mathrm{rot}$, whereas dLW and RV in both channels, VO~7436, CRX~NIR, and CCF~BIS~NIR do not show the highest peak but still significant power at or close to $f_\mathrm{rot}$. In the case of RV~VIS, the strongest peak is separated from a weaker peak at $f_\mathrm{rot}$ by about $0.003\,\mathrm{d}^{-1}$, and the window function suggests that it is a yearly alias. The signals in dLW, CRX, RV, and CCF~FWHM in the VIS channel were also reported by \citet{2021A&A...652A..28L}. We note that dLW, CRX, CCF~FWHM, and CCF~BIS show more significant signals in the NIR channel than in the VIS channel. This is unexpected because the NIR wavelength range contains substantially fewer spectroscopic features \citep{2018A&A...612A..49R}, resulting in larger statistical fluctuations of the NIR parameters. A possible explanation is an asymmetric magnetic topology, which would result in larger variations of the NIR indicators because Zeeman broadening is stronger at larger wavelengths.

As for the full dataset, we calculated the mean values, standard deviations, and $\log p(f_\mathrm{rot})$ from the GLS periodogram of each indicator for two data subsets, and we show the results in Table~\ref{table:Ross318}. Several indicators that show significant power at $f_\mathrm{rot}$ in the full dataset also show significant power in one or both subsets, with the second subset containing more data points and more significant signals in more indicators. The $\log p(f_\mathrm{rot})$ values are generally higher in the full dataset than in either subset, implying that the signals are rather stable and become stronger as more data points are considered.

\subsection{TYC~3529-1437-1}
\label{subsection.TYC}
For the moderately active M2.0\,V star \object{TYC~3529-1437-1}, with a rotation period of $15.8\,$d, we show the GLS periodograms in Fig.~\ref{fig:TYC_gls}. H$\alpha$, two \ion{Ca}{ii}~IRT lines, and RV~NIR all show the most significant signal at $f_\mathrm{rot}$ or its daily alias frequency. \ion{Na}{i}~D, \ion{Ca}{ii} IRT-a, and RV~VIS also show significant power, but not the highest periodogram peak at the rotation frequency. For CRX~VIS and RV~VIS, the most significant peak instead appears at its second harmonic $2\,f_\mathrm{rot}$ or the yearly alias thereof. We therefore tabulate not only the mean values, standard deviations, and $\log p(f_\mathrm{rot})$, but also $\log p(2\,f_\mathrm{rot})$ in Table~\ref{table:TYC}. \citet{2021A&A...652A..28L} reported the same signals and an additional weak signal in CCF~BIS~VIS that does not fulfil our $3\sigma$ significance criterion.

The same quantities as for the full dataset are also tabulated for two data subsets in Table~\ref{table:TYC}. In the first subset, both \ion{He}{i} lines, H$\alpha$, all three \ion{Ca}{ii}~IRT lines, TiO~7050, VO~7436, CRX~VIS, RV~VIS, and CCF~FWHM~NIR all show significant power at $f_\mathrm{rot}$, whereas no indicator shows significant power at $2\,f_\mathrm{rot}$. In contrast, in the second subset, TiO~7050, CRX~VIS, RV~VIS, and RV~NIR show significant power at $2\,f_\mathrm{rot}$, but only dLW~NIR shows significant power at $f_\mathrm{rot}$. The signals at $f_\mathrm{rot}$ in the first subset and $2\,f_\mathrm{rot}$ in the second subset are particularly strong in TiO~7050 and RV~VIS, but their presence in only one subset leads to less significant or insignificant signals in the full dataset.

We also find that the standard deviations of most indicators are larger in the second subset, and the pEW$'$ values of chromospheric lines and the indices of photosperic bands are lower, indicating stronger chromospheric emission and stronger magnetic fields or a higher spot coverage in the photosphere. The stronger chromospheric emission can explain why the chromospheric indicators show no significant signals in the second subset, similarly to the more active YZ~CMi not showing any significant signals in the chromospheric indicators as described in Sect.~\ref{subsection.YZCMi}. A higher number of active regions in the photosphere leads more likely to a more symmetric distribution over the hemispheres that can cause the signals at $2\,f_\mathrm{rot}$, similarly to more symmetric star-spot distributions being able to cause two dips per rotation in photometric light curves, while an asymmetric distribution causes one dip \citep[e.g.][]{2020ApJ...901...14B}. We explore this similarity further in Section~\ref{section.phot}.

\subsection{EV~Lac}
\label{subsection.EVLac}
The periodicities in activity indicators of the very active M3.5\,V star \object{EV~Lac} and their relation to the rotation period of $4.38\,$d were studied in great detail by \citet{2022arXiv220300415J}. We show the GLS periodograms of all our indicators in Fig.~\ref{fig:EVLac_gls} and tabulate the mean values, standard deviations, and probabilities $\log p(f_\mathrm{rot})$, $\log p(2\,f_\mathrm{rot})$, and $\log p(3\,f_\mathrm{rot})$ in Table~\ref{table:EVLac}. We find that dLW and the CCF~contrast in both channels, H$\alpha$, two of the \ion{Ca}{ii}~IRT lines, both TiO band indices, and CCF~FWHM~NIR show a significant peak at $f_\mathrm{rot}$, and in several cases weaker but still significant peaks at the higher harmonics. All three \ion{Ca}{ii}~IRT lines actually show the most significant peaks at $f_\mathrm{rot}+0.03\,\mathrm{d}^{-1}$, which appear to be alias signals caused by the observation sampling. In contrast, CRX, RV, and CCF~BIS in both channels, both \ion{He}{i} lines, and CCF~FWHM~VIS show the most significant signal at $2\,f_\mathrm{rot}$ and less significant or insignificant power at $f_\mathrm{rot}$ or $3\,f_\mathrm{rot}$.

We note that the indicators that predominantly show the rotation frequency are equally affected by all features on the hemisphere facing the observer (i.e. they depend on the `strength' of all visible surface features). In contrast, most of the indicators that show mostly the second harmonic are affected in different ways by features on the part of the visible hemisphere that moves towards the observer, and on the part that moves away from the observer (i.e. they depend on the `positions' of the signatures of the surface feature in the spectrum). This may explain the different behaviour of the indicators. However, the \ion{He}{i} lines differ from the other chromospheric lines, so it is also possible that different distributions of the features tracked by each indicator cause their different behaviour. For further discussion including correlations between the indicators, we refer the reader to \citet{2022arXiv220300415J}. The broader periodicity analysis by \citet{2021A&A...652A..28L} found the same signals in the CARMENES VIS data and presented EV~Lac as an example of a star that shows signals at multiple frequencies related to the rotation period.

In the analysis of the data subsets, for which the results are also tabulated in Table~\ref{table:EVLac}, EV~Lac shows the most complex behaviour of the four stars considered in this work. CRX, RV, and CCF~BIS in both spectrograph channels, \ion{He}{i}~$\lambda$10833\,{\AA}, dLW~VIS, and CCF~FWHM~VIS show significant signals at $2\,f_\mathrm{rot}$, whereas TiO~7050 shows the most significant signals at $f_\mathrm{rot}$ and its daily alias in the first subset. In the second subset, CRX, RV, and CCF~BIS in the VIS channel favour $f_\mathrm{rot}$, whereas \ion{He}{i}~$\lambda$10833\,{\AA} still favours $2\,f_\mathrm{rot}$, and the other indicators show no significant power at the rotation frequency or its second and third harmonics. Finally, in the third subset, CRX and RV in both channels, \ion{He}{i}~$\lambda$10833\,{\AA}, and CCF~FWHM and CCF~BIS in the VIS channel show the most significant signals at $2\,f_\mathrm{rot}$, whereas dLW and the CCF~contrast in both channels, TiO~7050, and the chromospheric indicators except for the \ion{He}{i} lines favour $f_\mathrm{rot}$. Our results agree with the results reported by \citet{2022arXiv220300415J} for their combined maps 6 and 7, which correspond to the first 26 data points in our third subset. They compared their results with a different subset of 26 randomly selected data points, for which they found similar results to the full dataset. In our larger first and third subsets, some indicators also show significant power at $3\,f_\mathrm{rot}$.

There are no general trends in the standard deviations or mean values between the subsets except for a slight decrease in H$\alpha$ emission. While this suggests a lower activity level in the third subset that could explain why most chromospheric indicators show significant rotation signals only in this subset, it remains unclear why different indicators favour different periods in the different subsets. We do not find similar behaviour for YZ~CMi in Sect.~\ref{subsection.YZCMi}, although both the normalised H$\alpha$ luminosity and the average magnetic fields of EV~Lac and YZ~CMi are similar \citep{2019A&A...626A..86S}. However, the Zeeman-Doppler images from \citet{2008MNRAS.390..567M} reveal a more complex magnetic topology for EV~Lac, which results in a more complex surface feature distribution for EV~Lac \citep[e.g.][]{2022arXiv220300415J} than for YZ~CMi \citep[e.g.][]{2020A&A...641A..69B}. This is likely the reason for the very different behaviour we find for these two stars.

\subsection{YZ~CMi}
\label{subsection.YZCMi}
\object{YZ~CMi} is one of the most active M dwarfs in the CARMENES sample, and with spectral type M4.5\,V and a rotation period of $2.78\,$d, it is both the latest type and the fastest rotating star considered in this work. All GLS periodograms for YZ~CMi are shown in Fig.~\ref{fig:YZCMi_gls}, and the mean values, standard deviations, and $\log p(f_\mathrm{rot})$ are given in Table~\ref{table:YZCMi}. CRX, RV, and the CCF parameters in both channels, the two TiO indices, VO~7436, and dLW~VIS all show a significant peak at $f_\mathrm{rot}$ that is also the highest peak in the periodogram except for CCF~contrast~in the NIR. YZ~CMi was therefore also the example of a star that shows clear signals at the rotation period in \citet{2021A&A...652A..28L}. As CRX and RV are strongly anti-correlated \citep{2020A&A...641A..69B} for this star, we expect both these indicators to show similar signals.

Another noteworthy point is that none of the chromospheric line indicators show any significant signals at all. This might be caused by flaring events that lead to sporadic, non-periodic variations in the chromospheric emission lines on timescales ranging from minutes up to several hours \citep[e.g.][]{2013ApJS..207...15K}, that is, significantly shorter than $P_\mathrm{rot}$, whereas the other indicators are less sensitive to flaring events \citep[e.g.][]{2009A&A...505..859Z}. While the 2$\sigma$ clipping removes the strongest flaring events from our analysis, minor events or the decay phase of major events might still be included. However, EV~Lac and TYC~3529-1437-1 still show rotational signals in the chromospheric indicators despite occasional flaring events. It is possible that flaring events occur more randomly on YZ~CMi than on EV~Lac, where they occur more often at specific rotation phases \citep{2020MNRAS.499.5047M}. Another possible explanation for the missing signals in chromospheric line indicators is that there may be more plage regions in the chromosphere than active regions in the photosphere. A higher number of chromospheric features can lead to a more homogeneous distribution of them and, thus, less rotational variation in the chromospheric lines. In a sample of 13 active mid-to-late-type M dwarfs, \citet{2022arXiv220301344M} also found no rotational modulation in H$\alpha$, except, curiously, for the most active star in their sample.

Similarly to Ross~318, the results for the two analysed data subsets given in Table~\ref{table:YZCMi} reveal that the indicators that show significant power at $f_\mathrm{rot}$ in the full dataset also show significant power in one or both subsets. As in the full dataset, there are no significant signals in the chromospheric line indicators. Again, the second subset contains more data points and generally more significant signals than the first subset. The second subset also shows larger standard deviations in most indicators and more negative pEW$'$ values for the chromospheric lines. This stronger emission in these lines indicates a slightly higher activity level, which can result in stronger rotational variations. We note that the second subset for YZ~CMi contains the spectra that \citet{2020A&A...641A..69B} used to find a strong anti-correlation between CRX and RV and to constrain star-spot parameters.

%-------------------------------------------------------------------------------------------------------------------------------------------
\section{Comparison with photometry}
\label{section.phot}
In this section, we compare our results from the CARMENES spectra with photometric light curves observed by TESS. As the TiO~7050 index showed strong rotational signals for all four stars, we focus on this photospheric activity indicator. The TiO~7050 index is temperature-sensitive and thus affected by star spots that also cause modulations in the photometric light curve. However, it can also be affected by rotational and magnetic broadening \citep{2019A&A...623A..44S}. We did not analyse the light curves of Ross~318 because even two adjacent sectors cover only one rotation period.

\begin{figure}
  \resizebox{\hsize}{!}{\includegraphics{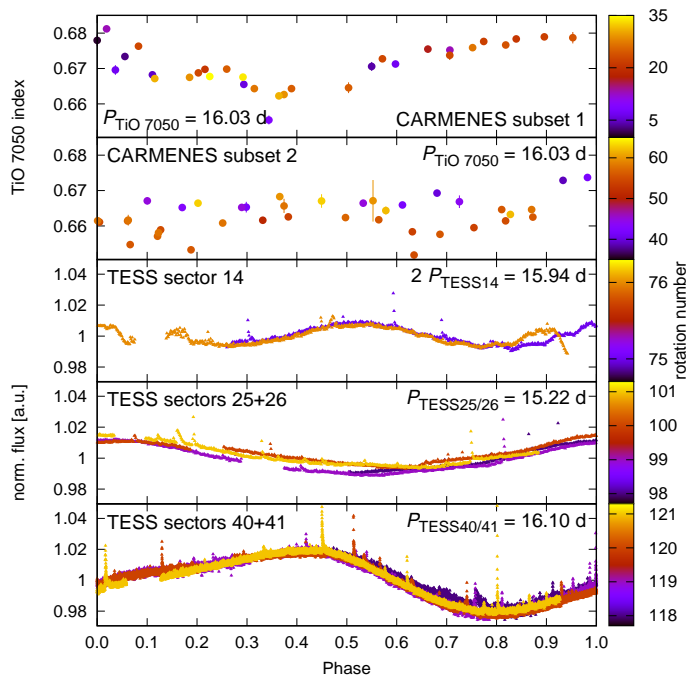}}
  \caption{Phase-folded time series of TiO~7050 index in CARMENES spectra and TESS light curves for TYC~3529-1437-1.}
  \label{fig:TYC_phasefold}
\end{figure}

Using GLS periodograms for TYC~3529-1437-1, we find the period $P_{\mathrm{TiO}~7050}=(16.03\pm 0.03)\,$d in the time series of the TiO~7050 index, and the periods $P_{\mathrm{TESS}14}=(7.97\pm 0.04)\,$d, $P_{\mathrm{TESS}25/26}=(15.22\pm 0.03)\,$d, and $P_{\mathrm{TESS}40/41}=(16.10\pm 0.03)\,$d in the TESS light curves from sector 14, sectors 25 and 26, and sectors 40 and 41, respectively. Although the formal uncertainties derived from the curvature of the GLS periogram peak are very small, only $P_{\mathrm{TESS}25/26}$ does not agree with either $P_\mathrm{rot}$ or $P_\mathrm{rot}/2$ within 3$\sigma$. In Fig.~\ref{fig:TYC_phasefold}, we show the TiO~7050 time series and the light curves phase-folded to the respective periods, or $2\,P_{\mathrm{TESS}14}$ in the case of TESS sector 14, using the barycentric Julian date 2457499.66246 of the first CARMENES observation as the epoch. The TiO~7050 time series shows one dip per rotation in the first subset of the CARMENES data, whereas the middle observations in the second subset show two dips and therefore cause the stronger signal at $2\,f_\mathrm{rot}$ than at $f_\mathrm{rot}$. The TESS light curves show two dips in sector 14, but only one dip in the later sectors.

\begin{figure}
  \resizebox{\hsize}{!}{\includegraphics{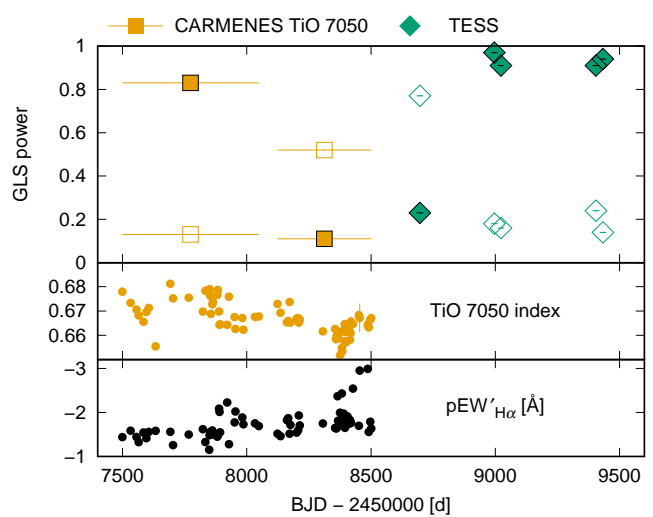}}
  \caption{Evolution of periodogram signals and activity indicators for TYC~3529-1437-1 over time. \textit{Top}: GLS periodogram power at $f_\mathrm{rot}$ (filled symbols) and $2\,f_\mathrm{rot}$ (open symbols) of the TiO~7050 index in the two CARMENES data subsets (orange boxes) and of TESS photometry in five TESS observation sectors (green diamonds) as a function of the barycentric Julian date. The bars reflect the time spanned by each set of data. \textit{Bottom}: Time series of TiO~7050 index (orange circles) and pEW$'_{\mathrm{H}\alpha}$ (black circles) derived from CARMENES spectra.}
  \label{fig:TYC_power}
\end{figure}

In the top panel of Fig.~\ref{fig:TYC_power}, we show the GLS power at the rotation frequency and its second harmonic for the TiO~7050 index in the two CARMENES data subsets and for the TESS light curves in each sector. As can be seen in the time series in the bottom panel, the TiO~7050 index decreased and the H$\alpha$ emission increased, indicating a higher activity level in the second subset, which shows more power at $2\,f_\mathrm{rot}$ than at $f_\mathrm{rot}$ in the GLS periodogram. The power at $2\,f_\mathrm{rot}$ is also higher in TESS sector 14 with two dips per rotation in the light curve, but lower again in the later sectors with only one dip. Alternating episodes of one dip and two dips per rotation period are common in light curves of late-type stars and do not necessarily reflect the global spot coverage and activity level. One dip means that one hemisphere appears darker than the other, so there may be spots on both hemispheres, but more on one than on the other. In contrast, two dips require a certain phase distribution of the spots, which can also occur with a lower spot coverage \citep{2020ApJ...901...14B}. We therefore cannot infer a lower activity level from the single dip per rotation in the later TESS sectors. If the activity decreased again, this would be consistent with an activity cycle of the order of a few years, as is plausible for an early M dwarf \citep[e.g.][]{2016A&A...595A..12S}, with the maximum occurring during or after the second subset of the CARMENES observations.

\begin{figure}
  \resizebox{\hsize}{!}{\includegraphics{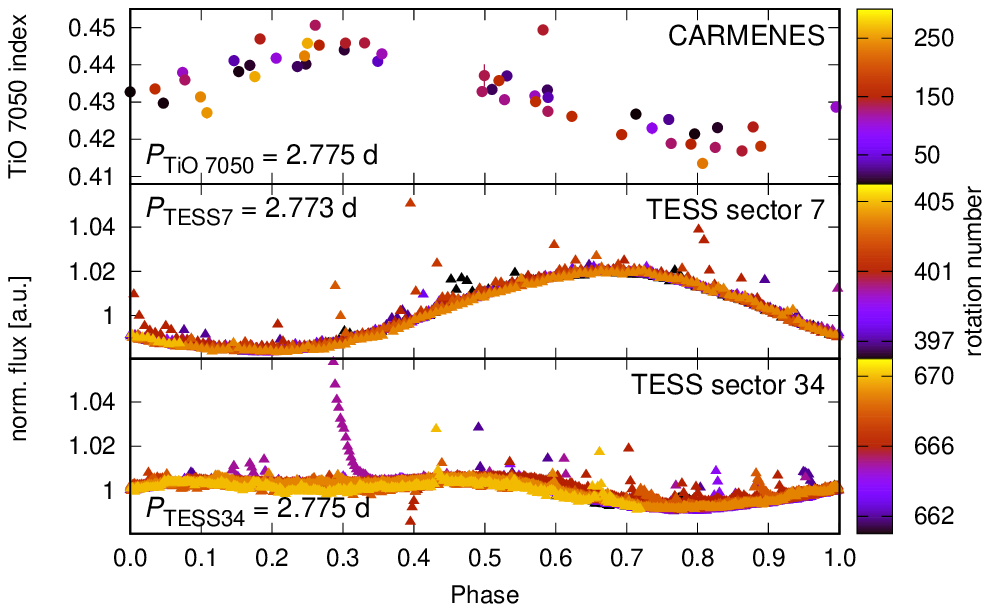}}
  \resizebox{\hsize}{!}{\includegraphics{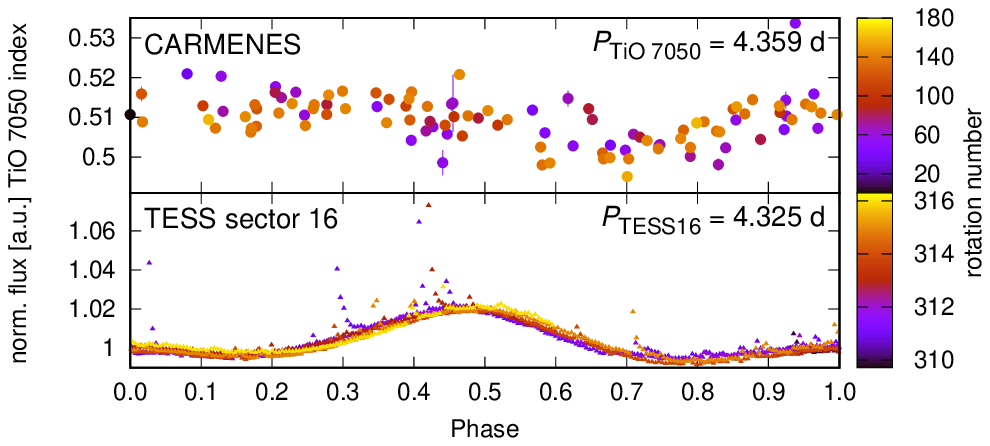}}
  \caption{Phase-folded time series of TiO~7050 index in CARMENES spectra and TESS light curves for YZ~CMi (\textit{top}) and EV~Lac (\textit{bottom}).}
  \label{fig:YZCMi_EVLac_phasefold}
\end{figure}

For YZ~CMi and EV~Lac, we show the phase-folded TiO~7050 time series and TESS light curves in Fig.~\ref{fig:YZCMi_EVLac_phasefold}. The epochs are again the barycentric Julian dates of the first CARMENES observations for each star (2457395.59368 for YZ~CMi and 2457398.35702 for EV~Lac), the periods are $P_{\mathrm{TiO}~7050}=(2.775\pm 0.001)\,$d, $P_{\mathrm{TESS}7}=(2.773\pm 0.001)\,$d, and $P_{\mathrm{TESS}34}=(2.775\pm 0.004)\,$d for YZ~CMi, and $P_{\mathrm{TiO}~7050}=(4.359\pm 0.004)\,$d and $P_{\mathrm{TESS}16}=(4.325\pm 0.017)\,$d for EV~Lac. All periods are in agreement with the respective $P_\mathrm{rot}$. YZ~CMi shows one dip per rotation in the TiO~7050 time series and the TESS sector 7 light curve and two dips in TESS sector 34. However, the second dip is considerably weaker and does not cause more power at the second harmonic than at the rotation frequency in the GLS periodogram. Given the long-term stability of active regions suggested by \citet{2020A&A...641A..69B} and the stable rotational signals found in the spectroscopic indicators, it is remarkable that we see hints at their evolution in the light curves. While EV~Lac also shows two dips per rotation in the TESS light curve, but more GLS power at $f_\mathrm{rot}$, in this case it is the second peak that is considerably weaker than the first peak. We note that EV~Lac showed only one dip per rotation period in the photometric data from automated surveys used by \citet{2019A&A...621A.126D}, and the long-term light curves from \citet{2017ARep...61..221A} include both episodes with one dip and episodes with two dips per rotation period for this star.

%-------------------------------------------------------------------------------------------------------------------------------------------
\section{Summary and conclusions}
\label{section.conclusions}
We investigated periodicities in the strength of eight chromospheric emission lines, four photospheric absorption bands, and in dLW, CRX, RV, and CCF parameters from the CARMENES VIS and NIR channels for four early-to-mid type M dwarfs with different activity levels. While the four stars were selected because they all showed clear rotational signals in more than two activity indicators in a previous activity analysis of the CARMENES sample \citep{2019A&A...623A..44S}, we find that they show the signals in different indicators and that these signals evolve in different ways.

\begin{figure}
  \resizebox{\hsize}{!}{\includegraphics{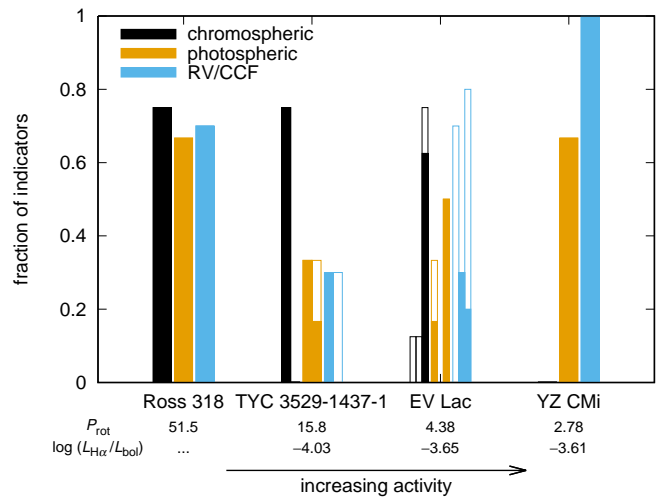}}
  \caption{Fractions of eight spectral lines with a chromospheric component, six photospheric indicators (band indices and dLW), and ten RV and CCF indicators that show a significant signal at $f_\mathrm{rot}$ (solid) or $2\,f_\mathrm{rot}$ (open) in the GLS periodograms. The bars for TYC~3529-1437-1 and EV~Lac are split into halves and thirds, respectively, corresponding to the analysed data subsets in chronological order.}
  \label{fig:results_histogram}
\end{figure}

The histogram in Fig.~\ref{fig:results_histogram} summarises how many chromospheric, photospheric, and CCF or RV-related indicators show a modulation with either the rotation frequency or its second harmonic. While the photospheric indicators and the CCF and RV-related indicators, particularly the TiO~7050 index and the RV itself, show rotational modulation at all activity levels, rotational signals in the chromospheric indicators are more common at lower activity levels. This is also true for observations of the same star at different activity levels and can be explained with an increasing flaring rate or an increasing number of plage regions leading to a more homogeneous distribution of active regions in the chromosphere with increasing global activity. Overall, our results agree with the findings of \citet{2021A&A...652A..28L} for a larger sample of M dwarfs.

Ross~318 and YZ~CMi, the least and the most active of the four studied stars, both show stable signals corresponding to $P_\mathrm{rot}$ throughout the CARMENES observations, whereas the other two stars exhibit changing signals over time. The moderately active star TYC~3529-1437-1 shows signals at the rotation frequency in several indicators during a less active episode, but signals at the second harmonic $2\,f_\mathrm{rot}$ during a more active episode. Our comparison with TESS photometry suggests that this corresponds to the common phenomenon of light curves showing one dip per rotation at certain times, but two dips per rotation at other times \citep[e.g.][]{2018ApJ...863..190B}. For the very active star EV~Lac, with its non-axisymmetric magnetic field \citep{2008MNRAS.390..567M}, we find a more complex behaviour with different indicators showing the strongest rotational signal at different frequencies during the same episode, and some indicators favouring different frequencies during other episodes in agreement with the results of \citet{2022arXiv220300415J}. Particularly for more active stars, it is therefore useful to also search for signals in activity indicators in parts of the dataset, not only in the full dataset.

%-------------------------------------------------------------------------------------------------------------------------------------------
\begin{acknowledgements}
CARMENES is an instrument for the Centro Astron\'omico Hispano-Alem\'an de Calar Alto (CAHA, Almer\'{\i}a, Spain). CARMENES is funded by the German Max-Planck-Gesellschaft (MPG), the Spanish Consejo Superior de Investigaciones Cient\'{\i}ficas (CSIC), the European Union through FEDER/ERF FICTS-2011-02 funds, and the members of the CARMENES Consortium (Max-Planck-Institut f\"ur Astronomie, Instituto de Astrof\'{\i}sica de Andaluc\'{\i}a, Landessternwarte K\"onigstuhl, Institut de Ci\`encies de l'Espai, Institut f\"ur Astrophysik G\"ottingen, Universidad Complutense de Madrid, Th\"uringer Landessternwarte Tautenburg, Instituto de Astrof\'{\i}sica de Canarias, Hamburger Sternwarte, Centro de Astrobiolog\'{\i}a and Centro Astron\'omico Hispano-Alem\'an),  with additional contributions by the Spanish Ministry of Science, the German Science Foundation through the Major Research Instrumentation Programme and DFG Research Unit FOR2544 ``Blue Planets around Red Stars'', the Klaus Tschira Stiftung, the states of Baden-W\"urttemberg and Niedersachsen, and by the Junta de Andaluc\'{\i}a.
S.V.J. also acknowledges the support of the DFG priority program SPP 1992 ``Exploring the Diversity of Extrasolar Planets (JE 701/5-1)''.
L.T.-O. also acknowledges support from the Israel Science Foundation (grant No. 848/16).
This paper includes data collected by the TESS mission, which are publicly available from the Mikulski Archive for Space Telescopes (MAST). Funding for the TESS mission is provided by NASA’s Science Mission directorate.
\end{acknowledgements}

\bibliographystyle{aa}
\bibliography{four_rotating_stars}

%-------------------------------------------------------------------------------------------------------------------------------------------
\begin{appendix}
\begin{figure*}
\section{Periodograms and tables}
\label{section.periodograms}
  \resizebox{\hsize}{!}{\includegraphics{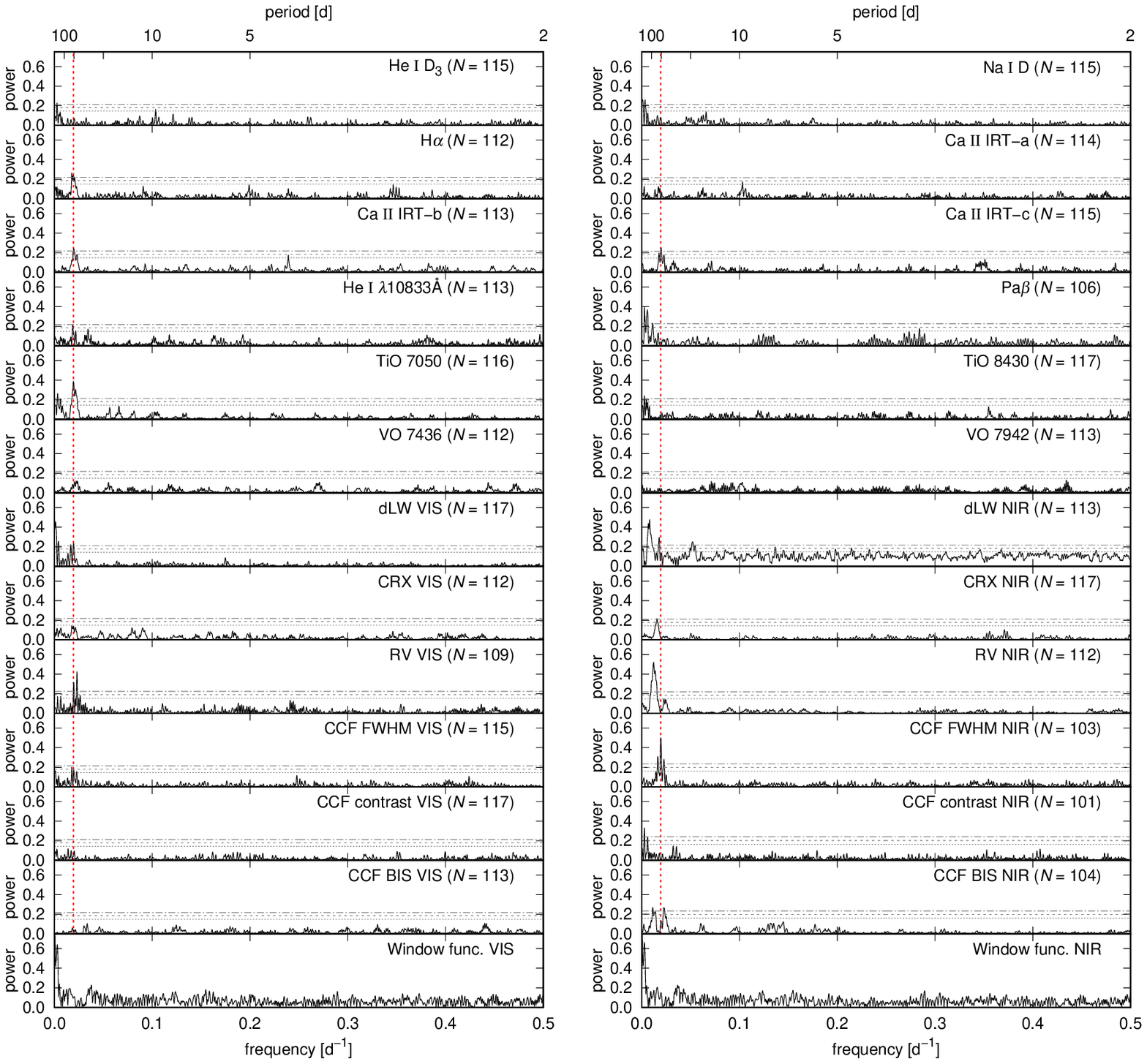}}
  \caption{GLS periodograms of our indicators and window functions for Ross~318. The indicators are the pEW$'$ of chromospheric emission lines (\ion{He}{i}~D$_3$, \ion{Na}{i}~D, H$\alpha$, and \ion{Ca}{ii} IRT in the visible-light channel of CARMENES, \ion{He}{i}~$\lambda$10833\,{\AA} and Pa$\beta$ lines in the near-infrared channel), indices of photospheric absorption bands (TiO~7050, TiO~8430, VO~7436, VO~7942), and differential line width (dLW), chromatic index (CRX), radial velocity (RV), CCF FWHM, CCF contrast, and CCF bisector inverse slope (BIS) in each channel. Horizontal lines indicate the analytical 10\% (dotted), 1\% (dashed), and 0.1\% (dash-dotted) false-alarm probability levels. $N$ is the number of used data points after a 2$\sigma$ clipping for each indicator. For most signals at a frequency $f$, there are aliases at $1\,\mathrm{d}^{-1} - f$. The red dotted line marks the rotation frequency $f_\mathrm{rot} = 0.0194\,\mathrm{d}^{-1}$ ($P_\mathrm{rot} = 51.5\,$d).}
  \label{fig:Ross318_gls}
\end{figure*}

\begin{sidewaystable*}
\caption{Mean values, standard deviations, and $\log p(f_\mathrm{rot})$ values for all indicators of Ross~318 in the full dataset and in two subsets.}
\label{table:Ross318}
\centering
\begin{tabular}{@{\extracolsep{6pt}}l ccc ccc ccc@{}}
\hline\hline
\noalign{\smallskip}
 & \multicolumn{3}{c}{full dataset ($N_\mathrm{obs}=120$)} & \multicolumn{3}{c}{January 2016 - January 2017 ($N_\mathrm{obs}=38$)} & \multicolumn{3}{c}{May 2017 - December 2017 ($N_\mathrm{obs}=82$)}\\
\cline{2-4} \cline{5-7} \cline{8-10}
\noalign{\smallskip}
Indicator & $\bar{x}$ & std($x$) & $\log p(f_\mathrm{rot})$ & $\bar{x}$ & std($x$) & $\log p(f_\mathrm{rot})$ & $\bar{x}$ & std($x$) & $\log p(f_\mathrm{rot})$\\
\hline
\noalign{\smallskip}
\ion{He}{i}~D$_3$ [{\AA}] & $0.071$ & $0.019$ & $-1.71$ & $0.076$ & $0.018$ & $-1.34$ & $0.068$ & $0.019$ & $-0.73$\\
\ion{Na}{i}~D [{\AA}] & $-1.27$ & $0.12$ & $-2.21$ & $-1.25$ & $0.12$ & $-1.35$ & $-1.27$ & $0.13$ & $-1.42$\\
H$\alpha$ [{\AA}] & $0.02$ & $0.06$ & $\mathbf{-7.14}$ & $0.05$ & $0.04$ & $-0.63$ & $0.01$ & $0.06$ & $\mathbf{-8.12}$\\
\ion{Ca}{ii}~IRT-a [{\AA}] & $0.085$ & $0.008$ & $\mathbf{-3.18}$ & $0.087$ & $0.007$ & $-2.14$ & $0.085$ & $0.009$ & $-2.26$\\
\ion{Ca}{ii}~IRT-b [{\AA}] & $0.128$ & $0.012$ & $\mathbf{-6.55}$ & $0.132$ & $0.011$ & $-1.49$ & $0.126$ & $0.013$ & $\mathbf{-6.69}$\\
\ion{Ca}{ii}~IRT-c [{\AA}] & $0.082$ & $0.008$ & $\mathbf{-6.86}$ & $0.085$ & $0.006$ & $-2.00$ & $0.081$ & $0.008$ & $\mathbf{-6.20}$\\
\ion{He}{i}~$\lambda$10833\,{\AA} [{\AA}] & $0.06$ & $0.09$ & $\mathbf{-5.83}$ & $0.050$ & $0.006$ & $\mathbf{-4.04}$ & $0.06$ & $0.10$ & $\mathbf{-3.44}$\\
Pa$\beta$ [{\AA}] & $-0.005$ & $0.013$ & $\mathbf{-3.17}$ & $-0.004$ & $0.004$ & $-2.45$ & $-0.006$ & $0.015$ & $-1.20$\\
TiO~7050 & $0.597$ & $0.005$ & $\mathbf{-11.91}$ & $0.596$ & $0.005$ & $-1.47$ & $0.597$ & $0.005$ & $\mathbf{-10.41}$\\
TiO~8430 & $0.831$ & $0.005$ & $-1.61$ & $0.831$ & $0.004$ & $-2.02$ & $0.831$ & $0.005$ & $-1.05$\\
VO~7436 & $0.935$ & $0.003$ & $\mathbf{-3.01}$ & $0.936$ & $0.003$ & $-1.17$ & $0.935$ & $0.003$ & $-2.71$\\
VO~7942 & $0.949$ & $0.003$ & $-1.13$ & $0.949$ & $0.003$ & $-1.75$ & $0.949$ & $0.003$ & $-0.89$\\
dLW~VIS [m$^2$\,s$^{-2}$] & $-1.0$ & $23.2$ & $\mathbf{-6.61}$ & $-7.1$ & $21.6$ & $-0.78$ & $1.8$ & $23.4$ & $\mathbf{-5.19}$\\
dLW~NIR [m$^2$\,s$^{-2}$] & $16.1$ & $53.7$ & $\mathbf{-8.27}$ & $42.9$ & $55.7$ & $\mathbf{-3.01}$ & $4.8$ & $48.6$ & $-1.78$\\
CRX~VIS [m\,s$^{-1}$\,Np$^{-1}$] & $0.4$ & $15.7$ & $\mathbf{-3.62}$ & $7.5$ & $13.8$ & $-2.75$ & $-2.8$ & $15.4$ & $\mathbf{-3.10}$\\
CRX~NIR [m\,s$^{-1}$\,Np$^{-1}$] & $-100.7$ & $1068.2$ & $\mathbf{-4.77}$ & $-21.9$ & $86.4$ & $-0.18$ & $-133.9$ & $1270.9$ & $\mathbf{-6.38}$\\
RV~VIS [m\,s$^{-1}$] & $-0.1$ & $2.6$ & $\mathbf{-8.68}$ & $0.7$ & $2.5$ & $\mathbf{-5.71}$ & $-0.4$ & $2.5$ & $\mathbf{-7.70}$\\
RV~NIR [m\,s$^{-1}$] & $29.7$ & $279.9$ & $\mathbf{-3.51}$ & $9.7$ & $19.0$ & $-1.14$ & $38.3$ & $334.1$ & $\mathbf{-3.53}$\\
CCF~FWHM~VIS [km\,s$^{-1}$] & $4.48$ & $0.02$ & $\mathbf{-5.38}$ & $4.47$ & $0.02$ & $-1.83$ & $4.48$ & $0.02$ & $\mathbf{-4.01}$\\
CCF~FWHM~NIR [km\,s$^{-1}$] & $5.81$ & $0.12$ & $\mathbf{-15.18}$ & $5.84$ & $0.07$ & $\mathbf{-4.21}$ & $5.80$ & $0.14$ & $\mathbf{-10.29}$\\
CCF~contrast~VIS [\%] & $23.3$ & $0.4$ & $-2.57$ & $23.3$ & $0.4$ & $-1.38$ & $23.3$ & $0.4$ & $-1.51$\\
CCF~contrast~NIR [\%] & $15.3$ & $0.4$ & $-1.62$ & $15.2$ & $0.4$ & $-2.87$ & $15.3$ & $0.4$ & $-0.35$\\
CCF~BIS~VIS [km\,s$^{-1}$] & $-0.007$ & $0.004$ & $-1.46$ & $-0.006$ & $0.005$ & $-0.27$ & $-0.008$ & $0.003$ & $-1.04$\\
CCF~BIS~NIR [km\,s$^{-1}$] & $0.08$ & $0.75$ & $\mathbf{-6.69}$ & $0.01$ & $0.03$ & $\mathbf{-3.69}$ & $0.11$ & $0.9$ & $\mathbf{-5.73}$\\
\hline
\end{tabular}
\tablefoot{$p(f)$ is the probability to obtain a power from Gaussian noise that is higher than the GLS power at frequency $f$. Boldfaced probability values indicate a significance of $3\sigma$ or higher.}
\end{sidewaystable*}

\begin{figure*}
  \resizebox{\hsize}{!}{\includegraphics{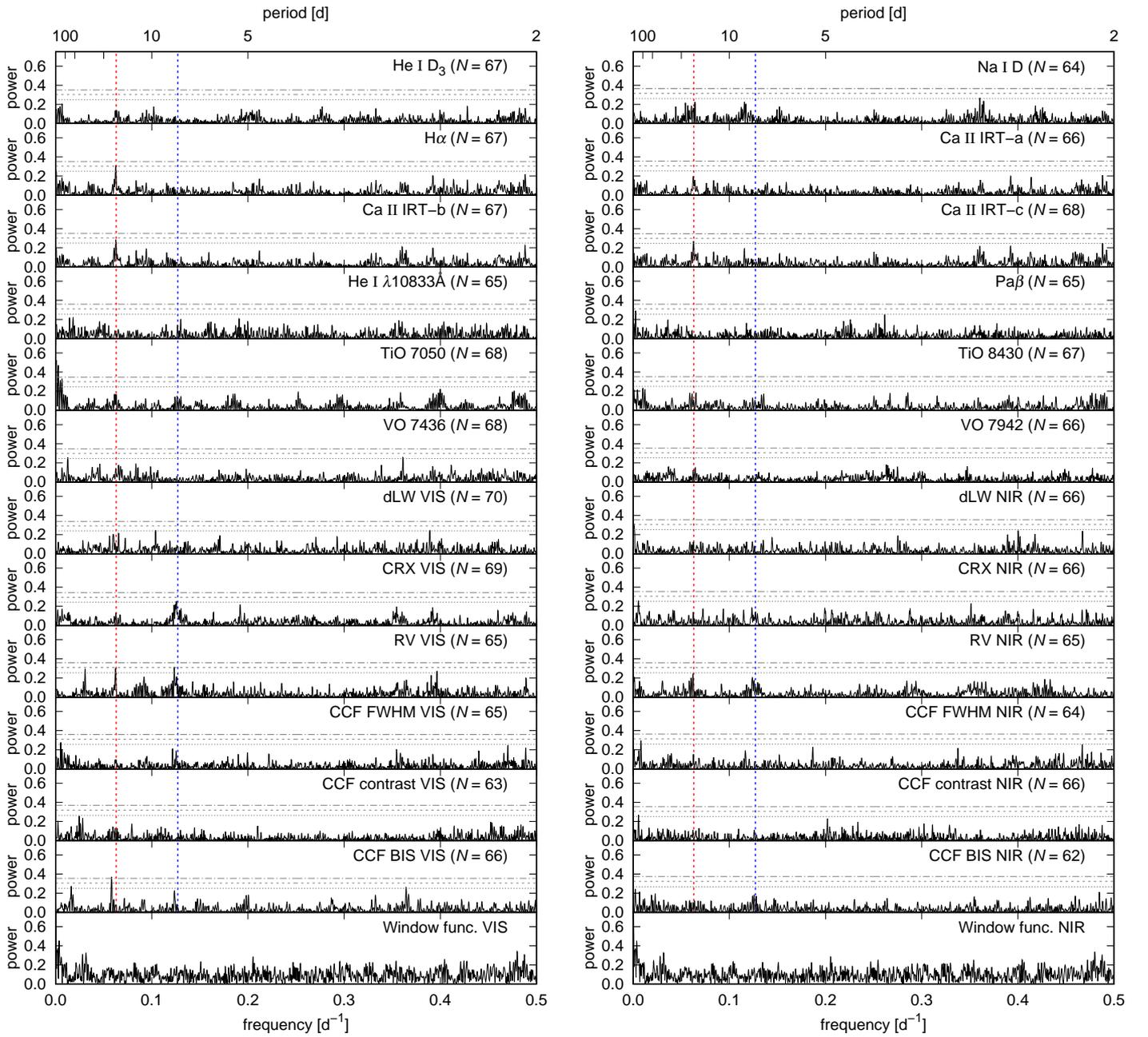}}
  \caption{Same as Fig.~\ref{fig:Ross318_gls}, but for TYC~3529-1437-1. The red dotted line marks the rotation frequency $f_\mathrm{rot} = 0.063\,\mathrm{d}^{-1}$ ($P_\mathrm{rot} = 15.8\,$d), while the blue dotted line marks its second harmonic $2\,f_\mathrm{rot}$.}
  \label{fig:TYC_gls}
\end{figure*}

\begin{sidewaystable*}
\caption{Mean values, standard deviations, and $\log p(f_\mathrm{rot})$ and $\log p(2\,f_\mathrm{rot})$ values for all indicators of TYC~3529-1437-1 in the full dataset and in two subsets.}
\label{table:TYC}
\small
\centering
\begin{tabular}{@{\extracolsep{6pt}}l cccc cccc cccc@{}}
\hline\hline
\noalign{\smallskip}
 & \multicolumn{4}{c}{full dataset ($N_\mathrm{obs}=71$)} & \multicolumn{4}{c}{April 2016 - October 2017 ($N_\mathrm{obs}=32$)} & \multicolumn{4}{c}{January 2018 - January 2019 ($N_\mathrm{obs}=39$)}\\
\cline{2-5} \cline{6-9} \cline{10-13}
\noalign{\smallskip}
Indicator & $\bar{x}$ & std($x$) & $\log p(f_\mathrm{rot})$ & $\log p(2\,f_\mathrm{rot})$ & $\bar{x}$ & std($x$) & $\log p(f_\mathrm{rot})$ & $\log p(2\,f_\mathrm{rot})$ & $\bar{x}$ & std($x$) & $\log p(f_\mathrm{rot})$ & $\log p(2\,f_\mathrm{rot})$\\
\hline
\noalign{\smallskip}
\ion{He}{i}~D$_3$ [{\AA}] & $-0.13$ & $0.05$ & $-2.01$ & $-1.08$ & $-0.12$ & $0.03$ & $\mathbf{-4.12}$ & $-1.29$ & $-0.15$ & $0.05$ & $-0.33$ & $-1.49$\\
\ion{Na}{i}~D [{\AA}] & $-0.34$ & $0.09$ & $\mathbf{-3.35}$ & $-1.24$ & $-0.32$ & $0.07$ & $-2.63$ & $-0.61$ & $-0.36$ & $0.10$ & $-2.03$ & $-0.83$\\
H$\alpha$ [{\AA}] & $-1.7$ & $0.4$ & $\mathbf{-5.06}$ & $-0.80$ & $-1.6$ & $0.3$ & $\mathbf{-4.75}$ & $-1.34$ & $-1.9$ & $0.4$ & $-0.92$ & $-0.47$\\
\ion{Ca}{ii}~IRT-a [{\AA}] & $-0.17$ & $0.04$ & $\mathbf{-2.91}$ & $-0.77$ & $-0.16$ & $0.04$ & $\mathbf{-3.85}$ & $-1.41$ & $-0.18$ & $0.05$ & $-0.36$ & $-1.46$\\
\ion{Ca}{ii}~IRT-b [{\AA}] & $-0.22$ & $0.05$ & $\mathbf{-4.61}$ & $-0.91$ & $-0.21$ & $0.04$ & $\mathbf{-5.10}$ & $-0.85$ & $-0.24$ & $0.05$ & $-1.01$ & $-1.52$\\
\ion{Ca}{ii}~IRT-c [{\AA}] & $-0.17$ & $0.04$ & $\mathbf{-4.43}$ & $-1.09$ & $-0.16$ & $0.03$ & $\mathbf{-4.44}$ & $-0.45$ & $-0.18$ & $0.05$ & $-1.02$ & $-1.26$\\
\ion{He}{i}~$\lambda$10833\,{\AA} [{\AA}] & $-0.001$ & $0.015$ & $-0.90$ & $-2.04$ & $0.004$ & $0.010$ & $\mathbf{-4.02}$ & $-0.84$ & $-0.006$ & $0.017$ & $-0.43$ & $-1.49$\\
Pa$\beta$ [{\AA}] & $0.024$ & $0.005$ & $-0.59$ & $-1.08$ & $0.024$ & $0.005$ & $-0.71$ & $-0.55$ & $0.023$ & $0.006$ & $-0.12$ & $-0.73$\\
TiO~7050 & $0.667$ & $0.007$ & $-2.56$ & $-2.23$ & $0.671$ & $0.006$ & $\mathbf{-10.75}$ & $-0.87$ & $0.663$ & $0.005$ & $-0.82$ & $\mathbf{-5.05}$\\
TiO~8430 & $0.822$ & $0.005$ & $-1.58$ & $-1.87$ & $0.824$ & $0.004$ & $-2.27$ & $-0.89$ & $0.820$ & $0.005$ & $-1.96$ & $-1.75$\\
VO~7436 & $0.93$ & $0.004$ & $-2.11$ & $-0.88$ & $0.931$ & $0.003$ & $\mathbf{-3.22}$ & $-1.41$ & $0.929$ & $0.004$ & $-0.28$ & $-0.46$\\
VO~7942 & $0.961$ & $0.003$ & $-2.21$ & $-1.17$ & $0.962$ & $0.003$ & $-2.17$ & $-0.42$ & $0.961$ & $0.003$ & $-2.23$ & $-0.98$\\
dLW~VIS [m$^2$\,s$^{-2}$] & $-2.9$ & $13.1$ & $-1.63$ & $-1.34$ & $-1.9$ & $9.1$ & $-1.44$ & $-0.82$ & $-3.7$ & $15.5$ & $-1.86$ & $-1.51$\\
dLW~NIR [m$^2$\,s$^{-2}$] & $26.7$ & $52.7$ & $-1.82$ & $-1.44$ & $6.8$ & $37.4$ & $-2.79$ & $-0.76$ & $42.1$ & $57.4$ & $\mathbf{-4.13}$ & $-0.43$\\
CRX~VIS [m\,s$^{-1}$\,Np$^{-1}$] & $-1.1$ & $47.0$ & $-2.00$ & $\mathbf{-4.18}$ & $5.3$ & $41.4$ & $\mathbf{-3.76}$ & $-0.37$ & $-6.4$ & $50.5$ & $-0.57$ & $\mathbf{-5.25}$\\
CRX~NIR [m\,s$^{-1}$\,Np$^{-1}$] & $7.4$ & $70.2$ & $-2.05$ & $-2.05$ & $-1.8$ & $77.0$ & $-1.69$ & $-1.66$ & $14.5$ & $63.6$ & $-1.11$ & $-1.85$\\
RV~VIS [m\,s$^{-1}$] & $-0.4$ & $11.3$ & $\mathbf{-4.80}$ & $\mathbf{-3.22}$ & $1.6$ & $11.5$ & $\mathbf{-7.37}$ & $-1.12$ & $-2.0$ & $10.8$ & $-0.89$ & $\mathbf{-8.42}$\\
RV~NIR [m\,s$^{-1}$] & $2.5$ & $17.8$ & $\mathbf{-3.75}$ & $-2.58$ & $8.5$ & $14.7$ & $-2.76$ & $-0.86$ & $-2.1$ & $18.6$ & $-1.51$ & $\mathbf{-4.19}$\\
CCF~FWHM~VIS [km\,s$^{-1}$] & $5.01$ & $0.03$ & $-1.35$ & $-2.84$ & $5.01$ & $0.03$ & $-2.56$ & $-2.47$ & $5.02$ & $0.03$ & $-1.78$ & $-1.57$\\
CCF~FWHM~NIR [km\,s$^{-1}$] & $7.92$ & $0.16$ & $-2.19$ & $-1.08$ & $7.85$ & $0.12$ & $\mathbf{-3.64}$ & $-1.38$ & $7.98$ & $0.16$ & $-0.86$ & $-0.18$\\
CCF~contrast~VIS [\%] & $19.7$ & $0.2$ & $-1.77$ & $-0.94$ & $19.7$ & $0.2$ & $-2.26$ & $-0.15$ & $19.8$ & $0.2$ & $-0.71$ & $-1.09$\\
CCF~contrast~NIR [\%] & $18.7$ & $0.9$ & $-1.60$ & $-1.63$ & $18.8$ & $0.5$ & $-2.42$ & $-1.27$ & $18.7$ & $1.1$ & $-0.41$ & $-0.46$\\
CCF~BIS~VIS [km\,s$^{-1}$] & $0.003$ & $0.010$ & $-1.11$ & $-1.85$ & $0.003$ & $0.007$ & $-0.34$ & $-0.87$ & $0.002$ & $0.010$ & $-0.66$ & $-2.26$\\
CCF~BIS~NIR [km\,s$^{-1}$] & $0.01$ & $0.06$ & $-1.62$ & $-2.57$ & $0.01$ & $0.05$ & $-0.42$ & $-1.94$ & $0.01$ & $0.06$ & $-1.59$ & $-1.37$\\
\hline
\end{tabular}
\tablefoot{$p(f)$ is the probability to obtain a power from Gaussian noise that is higher than the GLS power at frequency $f$. Boldfaced probability values indicate a significance of $3\sigma$ or higher.}
\end{sidewaystable*}

\begin{figure*}
  \resizebox{\hsize}{!}{\includegraphics{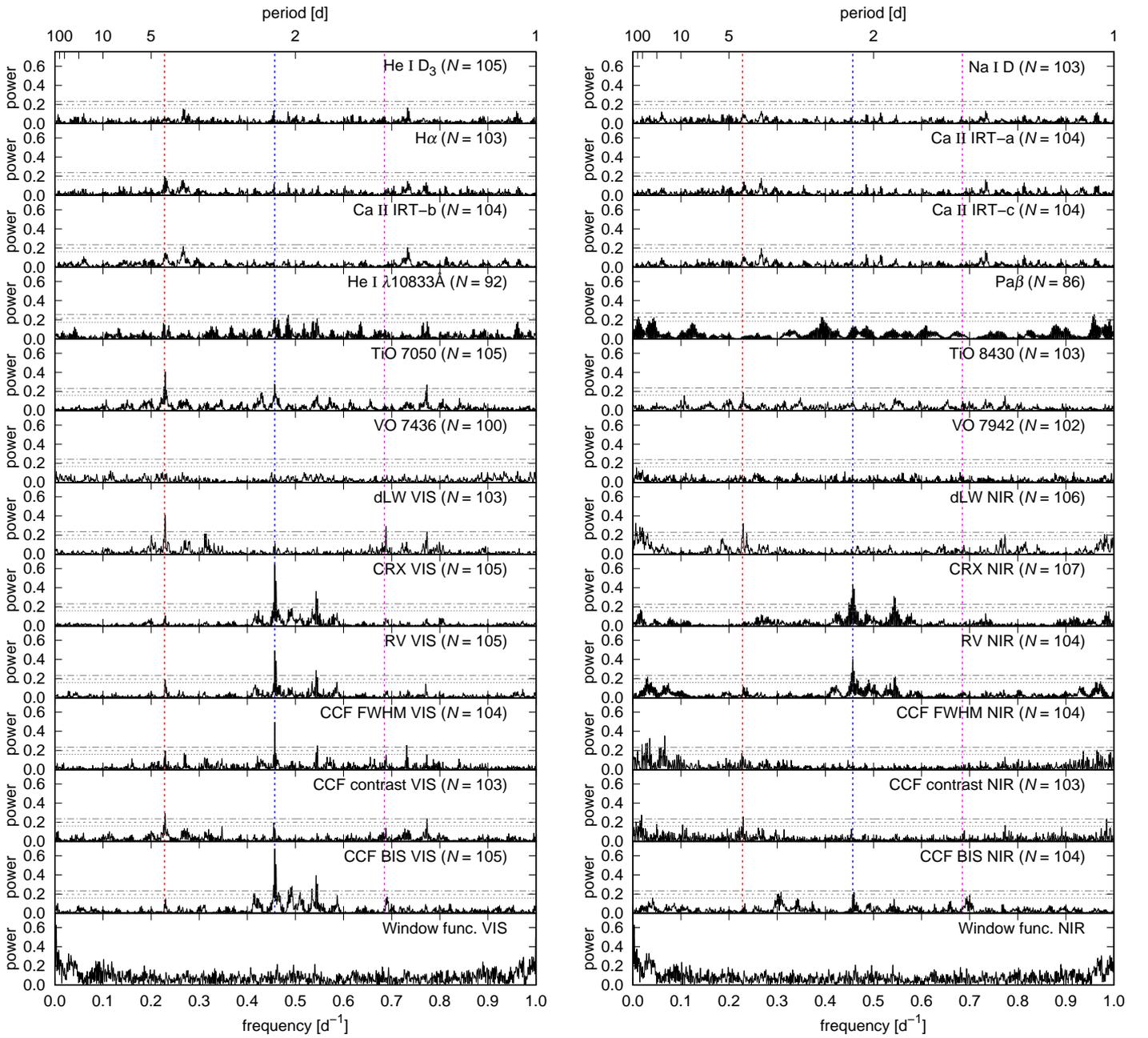}}
  \caption{Same as Fig.~\ref{fig:Ross318_gls}, but for EV~Lac. The red dotted line marks the rotation frequency $f_\mathrm{rot} = 0.228\,\mathrm{d}^{-1}$ ($P_\mathrm{rot} = 4.38\,$d), while the blue and magenta dotted lines mark its second and third harmonics $2\,f_\mathrm{rot}$ and $3\,f_\mathrm{rot}$, respectively.}
  \label{fig:EVLac_gls}
\end{figure*}

\begin{sidewaystable*}
\centering
\caption{Mean values, standard deviations, and $\log p(f_\mathrm{rot})$, $\log p(2\,f_\mathrm{rot})$, and $\log (3\,f_\mathrm{rot})$ values for all indicators of EV~Lac in the full dataset and in three subsets.}
\label{table:EVLac}
\begin{tabular}{@{\extracolsep{6pt}}l ccccc ccccc@{}}
\hline\hline
\noalign{\smallskip}
 & \multicolumn{5}{c}{full dataset ($N_\mathrm{obs}=107$)} & \multicolumn{5}{c}{June 2016 - January 2017 ($N_\mathrm{obs}=42$)}\\
\cline{2-6} \cline{7-11}
\noalign{\smallskip}
Indicator & $\bar{x}$ & std($x$) & $\log p(f_\mathrm{rot})$ & $\log p(2\,f_\mathrm{rot})$ & $\log p(3\,f_\mathrm{rot})$ & $\bar{x}$ & std($x$) & $\log p(f_\mathrm{rot})$ & $\log p(2\,f_\mathrm{rot})$ & $\log p(3\,f_\mathrm{rot})$\\
\hline
\noalign{\smallskip}
\ion{He}{i}~D$_3$ [{\AA}] & $-0.6$ & $0.4$ & $-1.53$ & $\mathbf{-3.22}$ & $-1.63$ & $-0.6$ & $0.5$ & $-0.20$ & $-1.41$ & $-1.61$\\
\ion{Na}{i}~D [{\AA}] & $-1.5$ & $0.5$ & $-2.41$ & $-1.43$ & $-1.55$ & $-1.5$ & $0.6$ & $-0.33$ & $-0.88$ & $-1.16$\\
H$\alpha$ [{\AA}] & $-5.5$ & $2.0$ & $\mathbf{-4.81}$ & $-2.66$ & $-1.61$ & $-5.8$ & $2.8$ & $-0.46$ & $-0.77$ & $-0.95$\\
\ion{Ca}{ii}~IRT-a [{\AA}] & $-0.20$ & $0.14$ & $\mathbf{-3.36}$ & $-1.86$ & $-1.75$ & $-0.20$ & $0.17$ & $-0.58$ & $-0.76$ & $-0.79$\\
\ion{Ca}{ii}~IRT-b [{\AA}] & $-0.27$ & $0.16$ & $\mathbf{-3.50}$ & $-1.19$ & $-1.28$ & $-0.29$ & $0.21$ & $-0.21$ & $-1.57$ & $-1.48$\\
\ion{Ca}{ii}~IRT-c [{\AA}] & $-0.16$ & $0.12$ & $-2.58$ & $-1.63$ & $-1.28$ & $-0.17$ & $0.16$ & $-0.15$ & $-1.10$ & $-1.22$\\
\ion{He}{i}~$\lambda$10833\,{\AA} [{\AA}] & $0.00$ & $0.04$ & $\mathbf{-3.45}$ & $\mathbf{-4.90}$ & $-1.99$ & $-0.01$ & $0.05$ & $-1.03$ & $\mathbf{-3.46}$ & $-2.34$\\
Pa$\beta$ [{\AA}] & $0.002$ & $0.012$ & $-0.53$ & $-2.52$ & $-1.53$ & $0.002$ & $0.016$ & $-0.06$ & $-2.08$ & $-1.42$\\
TiO~7050 & $0.509$ & $0.006$ & $\mathbf{-11.60}$ & $\mathbf{-7.15}$ & $-1.40$ & $0.510$ & $0.007$ & $\mathbf{-6.35}$ & $-1.80$ & $\mathbf{-3.19}$\\
TiO~8430 & $0.758$ & $0.005$ & $\mathbf{-4.54}$ & $-2.37$ & $-2.22$ & $0.758$ & $0.005$ & $-2.86$ & $-0.62$ & $-2.84$\\
VO~7436 & $0.867$ & $0.004$ & $-2.32$ & $-2.27$ & $-1.62$ & $0.866$ & $0.005$ & $-0.96$ & $-1.15$ & $-1.26$\\
VO~7942 & $0.944$ & $0.004$ & $-1.46$ & $-1.66$ & $-2.37$ & $0.944$ & $0.004$ & $-0.43$ & $-0.83$ & $-1.12$\\
dLW~VIS [m$^2$\,s$^{-2}$] & $3.8$ & $62.2$ & $\mathbf{-11.48}$ & $-2.59$ & $\mathbf{-7.46}$ & $5.0$ & $64.5$ & $-2.42$ & $\mathbf{-3.75}$ & $\mathbf{-3.46}$\\
dLW~NIR [m$^2$\,s$^{-2}$] & $38.2$ & $84.0$ & $\mathbf{-8.67}$ & $-1.89$ & $-2.28$ & $50.7$ & $70.3$ & $-1.46$ & $-1.84$ & $-1.55$\\
CRX~VIS [m\,s$^{-1}$\,Np$^{-1}$] & $-2.4$ & $160.3$ & $-2.51$ & $\mathbf{-22.93}$ & $-1.81$ & $-6.4$ & $181.9$ & $-0.69$ & $\mathbf{-11.38}$ & $-1.05$\\
CRX~NIR [m\,s$^{-1}$\,Np$^{-1}$] & $9.6$ & $193.7$ & $-1.02$ & $\mathbf{-13.00}$ & $-1.12$ & $-21.1$ & $169.1$ & $-0.53$ & $\mathbf{-7.61}$ & $-0.74$\\
RV~VIS [m\,s$^{-1}$] & $-1.0$ & $50.4$ & $\mathbf{-4.61}$ & $\mathbf{-14.97}$ & $-1.68$ & $1.5$ & $52.7$ & $-0.89$ & $\mathbf{-10.62}$ & $-0.63$\\
RV~NIR [m\,s$^{-1}$] & $-4.5$ & $42.2$ & $-2.65$ & $\mathbf{-11.41}$ & $-1.31$ & $-8.6$ & $39.6$ & $-1.24$ & $\mathbf{-6.13}$ & $-1.51$\\
CCF~FWHM~VIS [km\,s$^{-1}$] & $7.12$ & $0.07$ & $\mathbf{-4.86}$ & $\mathbf{-14.99}$ & $\mathbf{-3.57}$ & $7.13$ & $0.08$ & $-0.93$ & $\mathbf{-7.31}$ & $\mathbf{-3.02}$\\
CCF~FWHM~NIR [km\,s$^{-1}$] & $9.20$ & $0.16$ & $\mathbf{-4.16}$ & $-2.18$ & $-1.58$ & $9.19$ & $0.15$ & $-2.29$ & $-1.70$ & $-1.71$\\
CCF~contrast~VIS [\%] & $20.2$ & $0.4$ & $\mathbf{-7.54}$ & $\mathbf{-4.66}$ & $\mathbf{-3.21}$ & $20.2$ & $0.4$ & $-1.76$ & $-2.77$ & $-2.74$\\
CCF~contrast~NIR [\%] & $15.3$ & $0.3$ & $\mathbf{-6.36}$ & $\mathbf{-3.03}$ & $-2.56$ & $15.2$ & $0.3$ & $-1.44$ & $-0.65$ & $-1.97$\\
CCF~BIS~VIS [km\,s$^{-1}$] & $-0.01$ & $0.04$ & $\mathbf{-3.51}$ & $\mathbf{-24.64}$ & $\mathbf{-4.07}$ & $-0.02$ & $0.04$ & $-0.60$ & $\mathbf{-14.02}$ & $-1.37$\\
CCF~BIS~NIR [km\,s$^{-1}$] & $-0.02$ & $0.10$ & $-2.27$ & $\mathbf{-5.42}$ & $\mathbf{-4.15}$ & $0.01$ & $0.11$ & $-2.07$ & $\mathbf{-6.87}$ & $-2.42$\\
\hline
\end{tabular}
\end{sidewaystable*}

\begin{sidewaystable*}
\centering
\setcounter{table}{2}
\caption{continued.}
\begin{tabular}{@{\extracolsep{6pt}}l ccccc ccccc@{}}
\hline\hline
\noalign{\smallskip}
 & \multicolumn{5}{c}{April 2017 - August 2017 ($N_\mathrm{obs}=18$)} & \multicolumn{5}{c}{August 2017 - December 2017 ($N_\mathrm{obs}=46$)}\\
\cline{2-6} \cline{7-11}
\noalign{\smallskip}
Indicator & $\bar{x}$ & std($x$) & $\log p(f_\mathrm{rot})$ & $\log p(2\,f_\mathrm{rot})$ & $\log p(3\,f_\mathrm{rot})$ & $\bar{x}$ & std($x$) & $\log p(f_\mathrm{rot})$ & $\log p(2\,f_\mathrm{rot})$ & $\log p(3\,f_\mathrm{rot})$\\
\hline
\noalign{\smallskip}
\ion{He}{i}~D$_3$ [{\AA}] & $-0.6$ & $0.2$ & $-0.75$ & $-0.71$ & $-1.31$ & $-0.6$ & $0.2$ & $-2.64$ & $-1.92$ & $-0.74$\\
\ion{Na}{i}~D [{\AA}] & $-1.5$ & $0.3$ & $-0.99$ & $-0.63$ & $-0.49$ & $-1.5$ & $0.4$ & $\mathbf{-3.72}$ & $-0.72$ & $-0.85$\\
H$\alpha$ [{\AA}] & $-5.5$ & $1.2$ & $-0.37$ & $-1.11$ & $-0.33$ & $-5.3$ & $1.3$ & $\mathbf{-4.20}$ & $-1.32$ & $-1.51$\\
\ion{Ca}{ii}~IRT-a [{\AA}] & $-0.19$ & $0.09$ & $-1.95$ & $-0.52$ & $-0.70$ & $-0.20$ & $0.11$ & $\mathbf{-3.70}$ & $-0.45$ & $-1.00$\\
\ion{Ca}{ii}~IRT-b [{\AA}] & $-0.27$ & $0.09$ & $-1.16$ & $-0.43$ & $-0.80$ & $-0.26$ & $0.12$ & $\mathbf{-4.09}$ & $-0.64$ & $-0.65$\\
\ion{Ca}{ii}~IRT-c [{\AA}] & $-0.16$ & $0.07$ & $-1.37$ & $-0.51$ & $-0.55$ & $-0.16$ & $0.09$ & $\mathbf{-3.25}$ & $-0.54$ & $-0.41$\\
\ion{He}{i}~$\lambda$10833\,{\AA} [{\AA}] & $0.00$ & $0.02$ & $-0.54$ & $\mathbf{-3.03}$ & $-1.18$ & $0.00$ & $0.03$ & $-1.85$ & $\mathbf{-4.36}$ & $-0.67$\\
Pa$\beta$ [{\AA}] & $0.002$ & $0.008$ & $-0.16$ & $-0.19$ & $-1.23$ & $0.002$ & $0.007$ & $-0.72$ & $-1.19$ & $-0.19$\\
TiO~7050 & $0.510$ & $0.004$ & $-0.97$ & $-1.18$ & $-2.14$ & $0.509$ & $0.006$ & $\mathbf{-4.93}$ & $\mathbf{-4.22}$ & $-0.59$\\
TiO~8430 & $0.759$ & $0.004$ & $-1.19$ & $-0.27$ & $-0.87$ & $0.756$ & $0.005$ & $-2.24$ & $-1.83$ & $-0.98$\\
VO~7436 & $0.868$ & $0.003$ & $-1.35$ & $-2.69$ & $-1.26$ & $0.868$ & $0.003$ & $-1.48$ & $-0.65$ & $-1.29$\\
VO~7942 & $0.944$ & $0.002$ & $-0.15$ & $-0.94$ & $-1.21$ & $0.945$ & $0.003$ & $-1.37$ & $-0.98$ & $-1.16$\\
dLW~VIS [m$^2$\,s$^{-2}$] & $19.5$ & $49.8$ & $-2.76$ & $-1.58$ & $-1.69$ & $-2.0$ & $63.3$ & $\mathbf{-5.40}$ & $-0.83$ & $-2.33$\\
dLW~NIR [m$^2$\,s$^{-2}$] & $77.2$ & $96.3$ & $-1.22$ & $-2.48$ & $-1.39$ & $13.5$ & $82.3$ & $\mathbf{-3.50}$ & $-0.47$ & $-1.23$\\
CRX~VIS [m\,s$^{-1}$\,Np$^{-1}$] & $-7.0$ & $109.8$ & $\mathbf{-3.84}$ & $-2.87$ & $-1.84$ & $3.1$ & $157.4$ & $-0.72$ & $\mathbf{-12.35}$ & $-0.71$\\
CRX~NIR [m\,s$^{-1}$\,Np$^{-1}$] & $70.4$ & $324.4$ & $-0.97$ & $-0.91$ & $-1.15$ & $12.5$ & $132.2$ & $-0.69$ & $\mathbf{-5.61}$ & $-0.88$\\
RV~VIS [m\,s$^{-1}$] & $7.0$ & $39.5$ & $\mathbf{-4.08}$ & $-1.86$ & $-1.94$ & $-6.5$ & $51.9$ & $-2.83$ & $\mathbf{-6.28}$ & $-0.79$\\
RV~NIR [m\,s$^{-1}$] & $-9.3$ & $41.7$ & $-2.21$ & $-1.86$ & $-1.40$ & $-0.6$ & $42.9$ & $-1.19$ & $\mathbf{-4.33}$ & $-0.69$\\
CCF~FWHM~VIS [km\,s$^{-1}$] & $7.13$ & $0.06$ & $-1.32$ & $-2.43$ & $-1.91$ & $7.11$ & $0.06$ & $-2.26$ & $\mathbf{-7.17}$ & $-1.69$\\
CCF~FWHM~NIR [km\,s$^{-1}$] & $9.20$ & $0.19$ & $-0.56$ & $-1.70$ & $-2.45$ & $9.20$ & $0.14$ & $-1.90$ & $-1.39$ & $-0.51$\\
CCF~contrast~VIS [\%] & $20.2$ & $0.2$ & $-1.09$ & $-1.31$ & $-2.01$ & $20.3$ & $0.3$ & $\mathbf{-7.92}$ & $-2.47$ & $-1.10$\\
CCF~contrast~NIR [\%] & $15.3$ & $0.4$ & $-2.27$ & $-2.76$ & $-2.59$ & $15.4$ & $0.3$ & $\mathbf{-4.12}$ & $-0.91$ & $-0.67$\\
CCF~BIS~VIS [km\,s$^{-1}$] & $-0.02$ & $0.03$ & $\mathbf{-3.64}$ & $-2.39$ & $-3.34$ & $-0.01$ & $0.03$ & $-0.96$ & $\mathbf{-11.34}$ & $-1.86$\\
CCF~BIS~NIR [km\,s$^{-1}$] & $-0.05$ & $0.14$ & $-1.29$ & $-1.17$ & $-2.54$ & $-0.03$ & $0.07$ & $-0.48$ & $-1.60$ & $\mathbf{-3.28}$\\
\hline
\end{tabular}
\tablefoot{$p(f)$ is the probability to obtain a power from Gaussian noise that is higher than the GLS power at frequency $f$. Boldfaced probability values indicate a significance of $3\sigma$ or higher.}
\end{sidewaystable*}

\begin{figure*}
  \resizebox{\hsize}{!}{\includegraphics{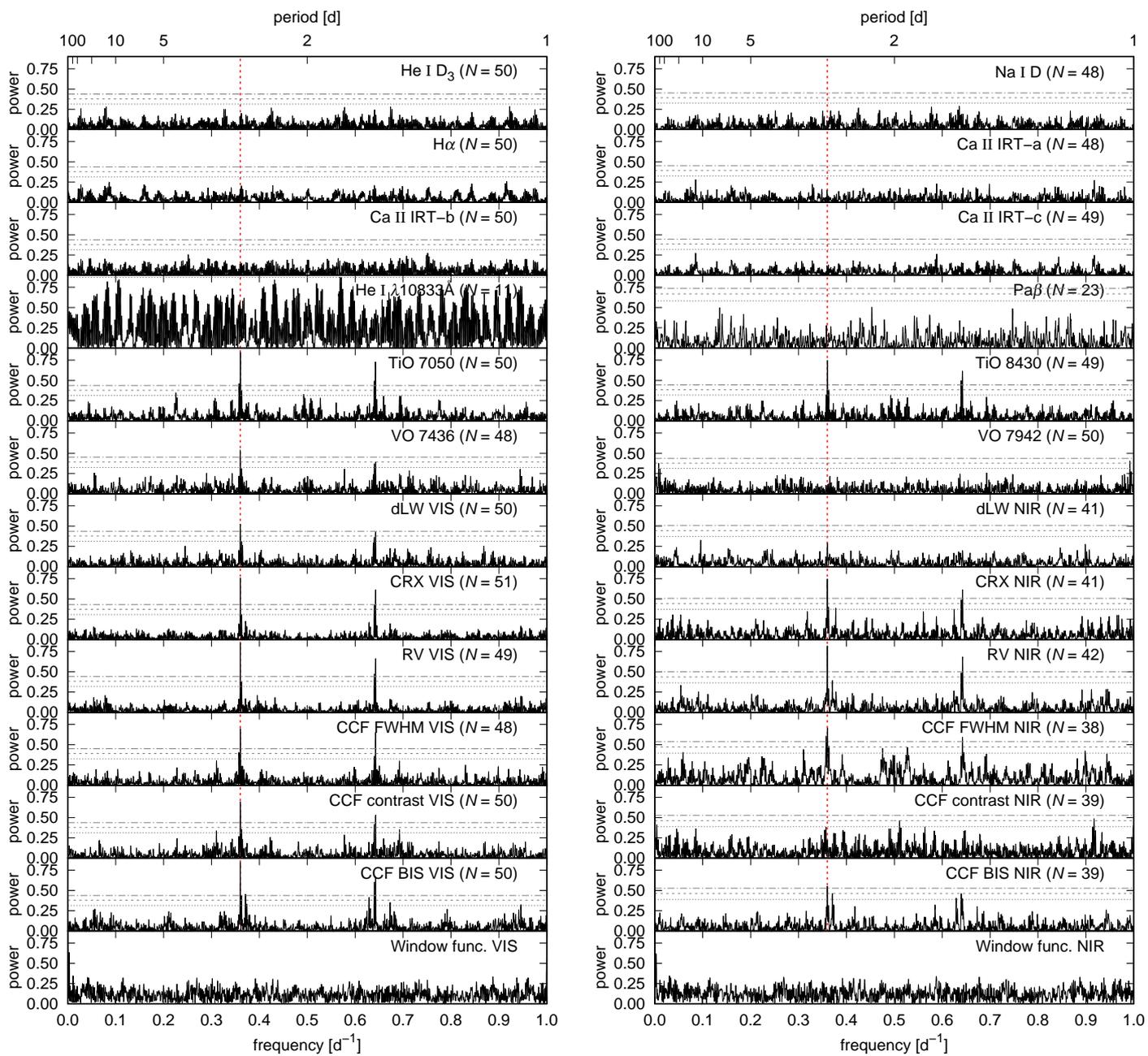}}
  \caption{Same as Fig.~\ref{fig:Ross318_gls}, but for YZ~CMi. The red dotted line marks the rotation frequency $f_\mathrm{rot} = 0.36\,\mathrm{d}^{-1}$ ($P_\mathrm{rot} = 2.78\,$d).}
  \label{fig:YZCMi_gls}
\end{figure*}

\begin{sidewaystable*}
\caption{Mean values, standard deviations, and $\log p(f_\mathrm{rot})$ values for all indicators of YZ~CMi in the full dataset and in two subsets.}
\label{table:YZCMi}
\centering
\begin{tabular}{@{\extracolsep{6pt}}l ccc ccc ccc@{}}
\hline\hline
\noalign{\smallskip}
 & \multicolumn{3}{c}{full dataset ($N_\mathrm{obs}=51$)} & \multicolumn{3}{c}{January 2016 - May 2016 ($N_\mathrm{obs}=17$)} & \multicolumn{3}{c}{September 2016 - May 2017 ($N_\mathrm{obs}=28$)}\\
\cline{2-4} \cline{5-7} \cline{8-10}
\noalign{\smallskip}
Indicator & $\bar{x}$ & std($x$) & $\log p(f_\mathrm{rot})$ & $\bar{x}$ & std($x$) & $\log p(f_\mathrm{rot})$ & $\bar{x}$ & std($x$) & $\log p(f_\mathrm{rot})$\\
\hline
\noalign{\smallskip}
\ion{He}{i}~D$_3$ [{\AA}] & $-0.9$ & $0.3$ & $-2.28$ & $-0.80$ & $0.12$ & $-2.20$ & $-0.9$ & $0.4$ & $-0.95$\\
\ion{Na}{i}~D [{\AA}] & $-1.6$ & $0.4$ & $-1.11$ & $-1.51$ & $0.16$ & $-0.51$ & $-1.6$ & $0.5$ & $-1.75$\\
H$\alpha$ [{\AA}] & $-7.5$ & $1.2$ & $-2.29$ & $-6.9$ & $0.6$ & $-0.82$ & $-7.8$ & $1.4$ & $-1.32$\\
\ion{Ca}{ii}~IRT-a [{\AA}] & $-0.19$ & $0.07$ & $-1.59$ & $-0.16$ & $0.04$ & $-0.30$ & $-0.20$ & $0.08$ & $-0.58$\\
\ion{Ca}{ii}~IRT-b [{\AA}] & $-0.22$ & $0.08$ & $-1.61$ & $-0.19$ & $0.05$ & $-1.03$ & $-0.24$ & $0.10$ & $-0.50$\\
\ion{Ca}{ii}~IRT-c [{\AA}] & $-0.10$ & $0.06$ & $-1.34$ & $-0.07$ & $0.03$ & $-1.16$ & $-0.11$ & $0.07$ & $-0.82$\\
\ion{He}{i}~$\lambda$10833\,{\AA} [{\AA}] & $-0.02$ & $0.09$ & $-1.45$ & \ldots & \ldots & \ldots & $-0.02$ & $0.10$ & $-0.68$\\
Pa$\beta$ [{\AA}] & $-0.006$ & $0.005$ & $-1.42$ & $-0.005$ & $0.003$ & $-0.47$ & $-0.006$ & $0.005$ & $-1.10$\\
TiO~7050 & $0.433$ & $0.010$ & $\mathbf{-19.30}$ & $0.434$ & $0.007$ & $\mathbf{-8.68}$ & $0.433$ & $0.011$ & $\mathbf{-11.26}$\\
TiO~8430 & $0.736$ & $0.006$ & $\mathbf{-13.84}$ & $0.736$ & $0.005$ & $\mathbf{-5.54}$ & $0.736$ & $0.007$ & $\mathbf{-10.38}$\\
VO~7436 & $0.849$ & $0.004$ & $\mathbf{-7.64}$ & $0.849$ & $0.004$ & $\mathbf{-3.95}$ & $0.849$ & $0.005$ & $\mathbf{-4.04}$\\
VO~7942 & $0.919$ & $0.004$ & $-1.29$ & $0.918$ & $0.003$ & $-1.33$ & $0.919$ & $0.004$ & $-1.02$\\
dLW~VIS [m$^2$\,s$^{-2}$] & $2.2$ & $60.6$ & $\mathbf{-7.63}$ & $-4.9$ & $54.9$ & $-2.81$ & $11.6$ & $62.1$ & $\mathbf{-4.24}$\\
dLW~NIR [m$^2$\,s$^{-2}$] & $30.3$ & $83.5$ & $-2.80$ & $13.3$ & $141.4$ & $-0.37$ & $40.4$ & $57.5$ & $-1.36$\\
CRX~VIS [m\,s$^{-1}$\,Np$^{-1}$] & $-34.8$ & $266.9$ & $\mathbf{-14.93}$ & $-38.1$ & $165.8$ & $\mathbf{-3.68}$ & $-81.2$ & $302.5$ & $\mathbf{-12.35}$\\
CRX~NIR [m\,s$^{-1}$\,Np$^{-1}$] & $-17.8$ & $163.9$ & $\mathbf{-11.43}$ & $56.8$ & $132.9$ & $-1.24$ & $-65.6$ & $167.3$ & $\mathbf{-8.92}$\\
RV~VIS [m\,s$^{-1}$] & $4.0$ & $87.6$ & $\mathbf{-19.29}$ & $5.2$ & $52.2$ & $\mathbf{-6.90}$ & $22.3$ & $97.2$ & $\mathbf{-15.35}$\\
RV~NIR [m\,s$^{-1}$] & $5.2$ & $60.0$ & $\mathbf{-14.60}$ & $-17.5$ & $43.0$ & $-1.19$ & $21.9$ & $61.6$ & $\mathbf{-15.40}$\\
CCF~FWHM~VIS [km\,s$^{-1}$]& $7.51$ & $0.10$ & $\mathbf{-11.77}$ & $7.50$ & $0.10$ & $\mathbf{-4.25}$ & $7.51$ & $0.11$ & $\mathbf{-6.55}$\\
CCF~FWHM~NIR [km\,s$^{-1}$] & $9.75$ & $0.13$ & $\mathbf{-9.18}$ & $9.75$ & $0.13$ & $-0.84$ & $9.75$ & $0.13$ & $\mathbf{-7.06}$\\
CCF~contrast~VIS [\%] & $19.8$ & $0.4$ & $\mathbf{-12.22}$ & $19.9$ & $0.3$ & $\mathbf{-4.67}$ & $19.7$ & $0.4$ & $\mathbf{-6.15}$\\
CCF~contrast~NIR [\%] & $16.6$ & $0.4$ & $\mathbf{-3.77}$ & $16.6$ & $0.7$ & $-0.91$ & $16.6$ & $0.3$ & $\mathbf{-2.91}$\\
CCF~BIS~VIS [km\,s$^{-1}$] & $-0.03$ & $0.06$ & $\mathbf{-19.23}$ & $-0.03$ & $0.05$ & $\mathbf{-3.97}$ & $-0.04$ & $0.07$ & $\mathbf{-14.97}$\\
CCF~BIS~NIR [km\,s$^{-1}$] & $0.01$ & $0.07$ & $\mathbf{-6.32}$ & $-0.01$ & $0.05$ & $-0.71$ & $0.01$ & $0.08$ & $\mathbf{-7.59}$\\
\hline
\end{tabular}
\tablefoot{$p(f)$ is the probability to obtain a power from Gaussian noise that is higher than the GLS power at frequency $f$. Boldfaced probability values indicate a significance of $3\sigma$ or higher.}
\end{sidewaystable*}
\end{appendix}
\end{document}